\documentclass[%
reprint,
superscriptaddress,
nofootinbib,
nolongbibliography,
 amsmath,amssymb,
 aps,
prd,
]{revtex4-1}

\usepackage{graphicx}%
\usepackage{dcolumn}%
\usepackage{bm}%
\usepackage{hyperref}%
\usepackage{xcolor}
\usepackage{soul}
\usepackage{algpseudocode}
\usepackage{algorithm}
\hypersetup{colorlinks, citecolor=blue, linkcolor=blue, urlcolor=blue}

\definecolor{lightred}{rgb}{1,0.5,0.5}
\definecolor{lightgreen}{rgb}{0.5,1,0.5}
\definecolor{lightblue}{rgb}{0.5,0.5,1}
\definecolor{lightcyan}{rgb}{0.5,0.75,0.75}
\definecolor{lightmagenta}{rgb}{0.75,0.5,0.75}
\definecolor{customgreen}{rgb}{0.494,1,0.502}

\newcommand{\be}{\begin{equation}}
\newcommand{\ee}{\end{equation}}
\newcommand{\beqa}{\begin{eqnarray}}
\newcommand{\eeqa}{\end{eqnarray}}
\newcommand{\dd}{{\rm d}}

\newcommand{\rank}{\text{rank}}

\newcommand{\norm}[1]{\left\lVert#1\right\rVert}

\newcommand{\ip}[2]{\langle #1, #2 \rangle}
\newcommand{\R}{\mathbb{R}}
\newcommand{\C}{\mathbb{C}}

\newcommand\y[1]{\mathcal{#1}} %
\newcommand\f[1]{\mathbf{#1}} %

\DeclareMathOperator{\argmin}{argmin}

\let\Re\relax
\DeclareMathOperator{\Re}{Re}
\DeclareMathOperator{\diag}{diag}

\begin{document}

\title{
Hankel low-rank matrix approximation for gravitational-wave data analysis
}

\author{Nicholas Geissler}
\affiliation{Department of Applied Mathematics and Statistics, Johns Hopkins University\\
3400 North Charles Street, Baltimore, MD 21218, USA}
\author{Vladimir Strokov}
\affiliation{Department of Physics and Astronomy, West Virginia University\\
135 Willey Street, P.O. Box 6315, Morgantown, WV 26506, USA}
\affiliation{Center for Gravitational Waves and Cosmology (GWAC), West Virginia University\\
Chestnut Ridge Research Building, Morgantown, WV 26505, USA}
\affiliation{William H. Miller III Department of Physics \& Astronomy, Johns Hopkins University\\
3400 North Charles Street, Baltimore, MD 21218, USA}
\thanks{Corresponding author}
\email{vladimir.strokov@mail.wvu.edu}
\author{Christian K\"ummerle}
\affiliation{Department of Mathematics, University of Central Florida\\ 
4000 Central Florida Blvd, Orlando, FL 32816, USA}
\author{Sergey Kushnarev}
\affiliation{Department of Applied Mathematics and Statistics, Johns Hopkins University\\
3400 North Charles Street, Baltimore, MD 21218, USA}
\author{Emanuele Berti}
\affiliation{William H. Miller III Department of Physics \& Astronomy, Johns Hopkins University\\
3400 North Charles Street, Baltimore, MD 21218, USA}
\email{berti@jhu.edu}

\begin{abstract}
Next-generation gravitational-wave (GW) detectors, such as the Laser Interferometer Space Antenna (LISA), will observe vast numbers of overlapping signals. Disentangling these signals from instrumental noise and from one another constitutes a significant data analysis challenge. We explore a denoising technique based on embedding time series into Hankel matrices: a superposition of $n$ (damped) sinusoids corresponds to a matrix of rank $2n$. Thus, the problem of signal extraction is reduced to a structured low-rank approximation problem. Using synthetic data tailored to GW applications, we benchmark three Hankel-based algorithms: ESPRIT, Cadzow iterations, and iteratively reweighted least squares (IRLS). Our test scenarios include isolated and multi-component monochromatic signals, the resolution of sources with closely spaced frequencies, and the recovery of black hole quasinormal modes (QNM). All three algorithms achieve near-optimal performance consistent with Fisher matrix bounds, evidenced by an inverse-square scaling of the mismatch with the signal-to-noise ratio. Furthermore, a proof-of-concept application to numerical relativity waveforms validates the ability of these algorithms to extract QNM frequencies from ringdown signals. Hankel low-rank approximation therefore offers a transparent, computationally efficient avenue for preprocessing GW time series.
\end{abstract}

\date{\today}

\maketitle

\section{Introduction\label{sec:intro}}

The past decade has been nothing short of revolutionary for gravitational-wave (GW) astronomy. Following the first observation of GWs from a binary black hole merger, GW150914~\cite{LIGOScientific:2016aoc}, the observing runs O1, O2, and O3 yielded $90$~confirmed detections of compact binary coalescences~\cite{LIGOScientific:2018mvr, LIGOScientific:2021usb,KAGRA:2021vkt}. The recently released version of the Gravitational-Wave Transient Catalog (GWTC-4.0) provides another 128~significant events discovered in the first half of the fourth observing run O4a~\cite{LIGOScientific:2025slb}. This remarkable success motivates the development of next-generation observatories. These include new ground-based facilities like Cosmic Explorer~\cite{Reitze:2019iox,Evans:2021gyd} and the Einstein Telescope~\cite{Punturo:2010zz,Maggiore:2019uih}, as well as space-based detectors such as the Laser Interferometer Space Antenna (LISA)~\cite{2017arXiv170200786A,Seoane:2021kkk,LISA:2022yao}, TianQin~\cite{TianQin:2015yph,Hu:2018yqb,TianQin:2020hid}, and Taiji~\cite{Hu:2017mde}, which will open the low-frequency GW window.

In contrast to traditional telescopes that can focus on specific regions of the sky, GW detectors are more akin to highly sensitive, omnidirectional microphones. Their raw data stream, typically from a few channels, is a mixture of noise and actual astrophysical signals. This naturally turns GW detectors into a testbed for both established and new signal processing techniques. One class of such techniques deals with denoising, i.e., the separation of the signals from noise. As part of a data processing pipeline, denoising can improve the signal-to-noise ratio (SNR) and facilitate the inference of the signal's parameters.

While neural networks have shown great promise for denoising GW signals~\cite{Shen:2017jkj,Torres-Forne:2018yvv,Shen:2019ohi,Wei:2019zlc,Mogushi:2021cpw,Kato:2022bxo,Bacon:2022lsm,Wang:2022quo,Murali:2022sba,Mao:2024jad,Xu:2024jlv,Lin:2024umj,Houba:2024tyn,Reissel:2025ykl}, the appeal of classical denoising techniques lies in their inherent transparency and well-established statistical properties. Examples of GW applications of these techniques include total variation methods~\cite{Torres:2014zoa,Torres:2015gaa,Torres-Forne:2016dwq,Torres-Forne:2016vjg,Barneo:2022ffq}, local polynomial approximation with relative intersection of confidence intervals~\cite{Lopac:2020xfd}, adaptive filtering~\cite{Chassande-Mottin:2000zoj,Shen:2024itu}, as well as wavelet-based methods~\cite{Klimenko:2004qh,Licciardi:2024bhv,Virtuoso:2024cmv}. With the next-generation observatories comes another challenge: their much larger detection volumes will inevitably contain overlapping signals. This gives rise to the global fit problem (also known as the ``cocktail party problem''), which is especially dramatic in LISA (see, e.g.,~\cite{Cornish:2005qw,Vallisneri:2008ye,MockLISADataChallengeTaskForce:2009wir,Littenberg:2010gf,Littenberg:2020bxy,Littenberg:2023xpl}), although ground-based detectors will also have to deal with it~\cite{Reali:2022aps,Reali:2023eug,Badaracco:2024alj,Katz:2024oqg,Tania:2025bsa}. In this context, a useful denoising technique should be effective at disentangling a superposition of signals.

In this paper, we explore denoising techniques that are based on the relation between time series and low-rank Hankel matrices. These methods have been extensively studied in the field of time series analysis (see, for example,~\cite{Gillard:2023hankel} and references therein), and their applications include time series denoising, completion, and forecasting. These techniques rely on two key facts: 
\begin{itemize}
    \item[(1)] A discrete time series~$h=\{h_l\}_{l=1}^{L}$ can be arranged into a matrix~$\f{H}$ with constant anti-diagonals, known as a Hankel matrix. 
    \item[(2)] If a time series~$\hat{h}=\{\hat{h}_l\}_{l=1}^{L}$ is a superposition of $n$ sinusoids (or more generally, damped sinusoids), the rank of its Hankel matrix~$\hat{\f{H}}$ is $r=2n$.
\end{itemize}

Thus, for a signal that is a combination of (damped) sinusoids, the problem of extracting the signal from noisy data reduces to finding a low-rank Hankel matrix~$\hat{\f{H}}$ which best approximates the original ``noisy'' Hankel matrix~$\f{H}$. Note that, although this method also works with exponentially growing signals, we restrict ourselves to either monochromatic or damped sinusoids, as these are characteristic of expected GW signals.

Here we test two low-rank Hankel denoising algorithms: Cadzow iterations~(see, e.g.,~\cite{Gillard:2023hankel}) and the iteratively reweighted least squares (IRLS) method~\cite{KM18_iTwist18,KuemmerleMayrinkVerdun-ICML2021}. For both algorithms, we run an array of numerical experiments on synthetic data containing a signal and white Gaussian noise. These toy signals are chosen to represent common use cases in GW data analysis:
\begin{enumerate}
\item[(1)] \textit{One monochromatic signal}. This experiment serves as a consistency check for the algorithms and it approximates a slowly evolving GW source. Important examples are double white dwarfs (DWDs), which are expected to be the most abundant class of LISA sources (see~\cite{LISA:2022yao} and references therein).
\item[(2)] \textit{A superposition of multiple monochromatic signals}. We test how well the algorithms perform on overlapping signals. An obvious GW application is the global fit problem. Another use case involves periodic signals that are not strictly monochromatic, such as a DWD source modulated by the LISA detector response. Such a signal is effectively a sum of discrete Fourier harmonics offset by multiples of $1\;\mbox{yr}^{-1}$~\cite{Strokov:2023ypy} (cf.\, appendix of~\cite{Cornish:2007if}). During denoising, these harmonics can be treated as independent monochromatic signals, while correlations between their amplitudes, frequencies, and phases can be determined in postprocessing.

This experiment includes two cases. In its simpler version, the number of components in a signal is known, while in a more realistic case that number is determined as part of the denoising pipeline. In the second case, we vary the target number of components to be denoised and estimate the residual. The experimental dependence of that residual on the target number of components~$n_{\rm trial}$ is then used to estimate the true value~$n_{\rm true}$.
\item[(3)] \textit{Frequency separation: two monochromatic signals with close frequencies}. In this experiment, we test how well the algorithms can resolve two signals with a small frequency separation. One of the GW applications is again the global fit problem, in which multiple signals may populate a narrow frequency band.
\item[(4)] \textit{Quasinormal modes (QNMs)}. The algorithms are tested on the ringdown part of a GW signal, which is approximately described as a superposition of damped sinusoids known as QNMs (see e.g.~\cite{Kokkotas:1999bd,Berti:2005ys,Berti:2009kk, Berti:2025hly}). To recover the QNM frequencies and damping times (equivalently, the complex frequencies) of a denoised ringdown signal, we use the ESPRIT (Estimation of Signal Parameters via Rotational Invariance Techniques) algorithm~\cite{RoyKailath_ESPRIT:1989}. This algorithm improves upon Prony methods~\cite{Eldar_2015} which were previously applied to ringdown signals~\cite{Berti:2007dg}.  
\end{enumerate}

More broadly, the ESPRIT algorithm can recover frequencies directly from noisy signals. For this reason, we use it as a baseline and compare it with the Cadzow and IRLS methods in our experiments. In the ESPRIT approach, finding the ``denoised'' signal is a two-stage process. Once the frequencies are known, the corresponding amplitudes and phases can be determined by solving a convex optimization problem (see Section~\ref{subsec:esprit}).

The remainder of this paper is organized as follows. In Section~\ref{sec:setup}, we describe our experimental setup, including the synthetic datasets and the mismatch metric used to assess denoising performance. In Section~\ref{sec:algo}, we provide the details of our methodology, namely, the embedding of the time series into a Hankel matrix, signal parameter estimation with ESPRIT, and the two iterative algorithms for the underlying structured low-rank matrix approximation: Cadzow iterations and IRLS. In Section~\ref{sec:results}, we present the denoising results across four scenarios: a monochromatic signal (Sec.~\ref{subsec:single}), a mixture of monochromatic signals (Sec.~\ref{subsec:multi}), frequency separation (Sec.~\ref{subsec:separation}), and QNMs (Sec.~\ref{subsec:quasinorm}). Finally, in Section~\ref{sec:discussion} we discuss the performance, strengths, and limitations of these algorithms and outline potential directions for future research.

\section{Experimental setup\label{sec:setup}}

In this section, we introduce the definition of a signal, describe our synthetic dataset, and explain how the mismatch is used to quantify the quality of denoising. Our full implementation of the experiment pipeline is publicly available online~\cite{lisahankel}.

A mixture of $n$~monochromatic signals can be represented in the following form: 
\begin{eqnarray}\label{eq:SumofExps}
    \hat{h}_l &=& \sum\limits_{k=1}^n \hat{h}_l^{(k)}, \\
    \hat{h}_l^{(k)} &=& a_k e^{-\gamma_k l}\sin(2\pi f_k l + \phi_k)\,,
\end{eqnarray}
where $f_k$, $a_k$, $\phi_k$, and $\gamma_k$ ($k=1,\ldots\,,n$) are the signals' frequencies, amplitudes, phases, and damping factors, respectively. Without loss of generality, we enumerate the subsequent timesteps with integer numbers $l=1,\ldots\,,L$\,. This effectively means that a timestep $\Delta t$ is absorbed into the unitless frequency $f_k=\overline{f}_k\Delta t$ and damping factor $\gamma_k=\Delta t/\tau_k$\,, where $\overline{f}_k$ and $\tau_k$ are the physical frequencies and damping times, respectively. In other words, one can always assume that the sampling rate is $1$~Hz (equivalently, the cadence is $1$~s) and rescale the results appropriately. In what follows we therefore use $\Delta t=1$~s for convenience. Note also that $f_k$ and $\gamma_k$ can be combined into the complex frequencies 
\be
\omega_k = 2\pi f_k + j\gamma_k\,,
\ee
where $j$ is the imaginary unit.

For numerical efficiency and reproducibility, we first produce two source datasets which are then used to generate data for specific experiments on the fly. The first dataset contains individual ($n=1$) monochromatic signals while the second contains noise realizations. There are $10,000$ time series of length $L=400$~s in each dataset. To put this length of a time series into context, released data for the most massive detected event GW231123~\cite{ligo:GW231123} is provided at a sampling rate of $4096$~Hz and, depending on the purpose, can be downsampled even further (see e.g.~\cite{Wang:2025rvn}). The event itself lasted $\sim 0.1$~s, which amounts to $\lesssim 1000$ data points. The scaling of these denoising algorithms with time-series length is discussed in Section~\ref{sec:discussion}.

Regarding parameters of the monochromatic signals, their frequencies follow a log-uniform distribution between $f_{\rm min}=2/T=5\times 10^{-3}$~Hz and $f_{\rm max}= 0.5/(2\Delta t)=0.25$~Hz, where $T=L\Delta t$ denotes the total observation time. Their amplitudes are also sampled from a log-uniform distribution between $a_{\rm min}=0.2$ and $a_{\rm max}=100$, while their phases are uniformly distributed in the range $[0,2\pi)$. The noise dataset provides random realizations of Gaussian white noise with the constant power spectral density (PSD) $S_{\rm n}=1$. A noise realization is a sequence of independent and identically distributed (i.i.d.) random variables $(\epsilon_1, \epsilon_2,\ldots\,, \epsilon_L)$ which are drawn from a normal distribution with zero mean and dispersion $\sigma_{\Delta t}=\sqrt{S_{\rm n}/2\Delta t}$ (see Appendix~\ref{app:noise} for details).

In experiments 1, 2, and 3, noisy data is generated on the fly. To this end, we randomly draw one or more signals from the signal source dataset and overlay them with 50~noise realizations which are drawn at random from the noise dataset such that the $l$-th sample of the noisy data is given by
\be
h_l = \sum\limits_{k=1}^{n}\hat{h}_l^{(k)} + \epsilon_l\,,
\ee
where $\epsilon_l$ $(l=1,\ldots\,, L)$ is a noise realization. For reproducibility, we record the indices of the signals and noise realizations. In the first experiment, only one signal is drawn from the source dataset, while for the second experiment, we consider signals with $n=3$, $n=5$, and $n=7$ and draw them independently before adding them up. In the case of frequency separation, we generate two monochromatic signals with unit amplitudes and random phases, and we set the relative difference in their frequencies $f_1$ and $f_2$ to be $\delta\in[0.01,0.25]$, such that $f_2 = (1 + \delta)f_1$. They are then added and overlaid with noise realizations to produce noisy data.

In experiment 4, we use a realistic gravitational waveform from the Simulating eXtreme Spacetimes (SXS) catalog~\cite{Scheel:2025jct,SXSPackage_v2025.0.17,SXSCatalogData_3.0.0} to serve as a proof-of-concept for QNM recovery with the techniques under consideration. A full systematic study of black hole spectroscopy is beyond the scope of this work; instead, we focus on a single, representative source, \texttt{SXS:BBH:0305}~\cite{Boyle:2019kee, SXS:BBH:0305}. This simulation describes the merger of a quasicircular binary black hole system with mass ratio $1.22$ and anti-aligned dimensionless spins $\chi_1=0.33$ and $\chi_2=-0.44$, resulting in a final remnant with mass $M_f = 0.9520\,M$ (in units of the total initial mass~$M$) and spin $\chi_f = 0.6921$. We analyze the Weyl scalar $\Psi_4$, using the plus polarization of the $(2,2)$ and $(3,2)$ spherical harmonic modes. For each trial, we consider the ringdown tail with a start time varying between the peak $t_{\rm peak}$ and $t_{\rm peak}+30M$.  In this experiment, we do not inject synthetic noise, but rather compare the performance of ESPRIT applied directly to the raw data against ESPRIT applied after Cadzow denoising. In the second case, we set the convergence tolerance to the double floating-point precision $10^{-16}$ to mitigate any numerical noise. 

In order to evaluate the quality of denoising, we use the mismatch
\be \label{eq:mismatch}
\mathcal{M} = 1 - \frac{(h\vert \hat{h})}{\sqrt{(h\vert h)(\hat{h}\vert \hat{h})}}\,,
\ee
where the inner product is defined as 
\be
\left(a\vert b\right) = 4\Re\int\limits_0^{+\infty}{\frac{a_F b_F^\star}{S_{\rm n}}\dd{F}}\,,
\ee
with $a_F$ and $b_F$ being the Fourier transforms of the respective signals. As usual, the (squared) SNR is given by the inner product of a waveform $h$ with itself: 
\be
\rho^2\equiv  (h\vert h) = 4\int\limits_0^{+\infty}{\frac{|h_F|^2}{S_{\rm n}}\dd{F}}.
\ee
The mismatch is a convenient quality metric, as it provides a single number which, in our case, is closely related to the best possible measurement uncertainties that follow from Fisher matrix analysis. Indeed, the Fisher matrix $\f{F}$ for parameters $\boldsymbol{\theta}$ of a waveform $h=h(\boldsymbol{\theta})$ arises in the linear approximation with respect to the parameters:
\be
\label{eq:FM-expansion}
(\Delta h\vert \Delta h) \approx (\Delta\boldsymbol{\theta})^{\top}\f{F}(\Delta \boldsymbol{\theta})\,,
\ee
\be
\Delta h \equiv \hat{h}(\boldsymbol{\theta} + \Delta\boldsymbol{\theta}) - \hat{h}(\boldsymbol{\theta})\,.
\ee

Now, the uncertainty $\Delta\boldsymbol{\theta}$ in the parameters of a denoising result depends on the noise realization. If that uncertainty is close to that predicted by a Fisher matrix analysis, the average of the right-hand side of Eq.~(\ref{eq:FM-expansion}) over noise realizations is simply
\beqa
\left\langle(\Delta\boldsymbol{\theta})^{\top}\f{F}(\Delta \boldsymbol{\theta})\right\rangle &=& F_{ab}\langle\Delta\theta_a\Delta\theta_b\rangle \nonumber \\
&=&F_{ab}\Sigma_{ab} = \delta_{aa} = \chi\,,
\eeqa
where $\f{\Sigma}$ is the covariance matrix, $\chi$ is the number of parameters, and we have used the fact that $\f{\Sigma} = \f{F}^{-1}$. On the other hand,
\be
(\Delta h\vert \Delta h)\approx 2\rho^2 - 2\rho^2 (1-\mathcal{M})\,,
\ee
whence (see also~\cite{McWilliams:2010eq,Lindblom:2008cm})
\be
\label{eq:fisher_prediction}
\mathcal{M}\approx \frac 12\frac{\chi}{\rho^2}\propto \rho^{-2}\,.
\ee
Therefore, for a properly working denoising routine, we expect this inverse square dependence. Note also that, for $n$ signals with the same template, this dependence---including the prefactor---stays the same if the SNR is normalized by the number of signals as
\be
\label{eq:normalized_snr}
\overline{\rho}\equiv \frac{\rho}{\sqrt{n}}.
\ee
In particular, if the signals do not overlap in the frequency domain, $\overline{\rho}$ coincides with the root mean square of individual SNRs.

We now proceed to describe the details of the algorithms we apply for signal denoising and frequency extraction.

\section{Algorithms\label{sec:algo}}

A time series, which can be understood as a vector $\f{h} = (h_1,h_2,\ldots\,, h_L) \in \C^{L}$, can be embedded into a $(d_1 \times d_2)$-dimensional so-called \emph{Hankel matrix}
\beqa
\label{eq:Hankel:def}
\f{H} = 
\begin{bmatrix}
h_1 & h_2 & \dots & h_{d_2} \\
h_2 & h_3 & \dots & h_{d_2+1} \\
\vdots & \vdots & \ddots & \vdots \\
h_{d_1} & h_{d_1+1} & \dots & h_{d_1+d_2-1}
\end{bmatrix}\,,
\eeqa
whose window length~(or number of rows) $d_1$ is a free parameter, which together with the number of columns~$d_2$ satisfies the relation $d_1+d_2-1=L$. The object \eqref{eq:Hankel:def} is at the core of our methodology due to the following observation: if the entries of $\f{H}$ satisfy
\begin{equation}
\label{eq:timeseries:model}
h_l = \hat{h}_l + w_l = \sum_{k=1}^{r} a_k e^{j\omega_k l} + w_l
\end{equation}
for a set of complex amplitudes $\{a_k\}_{k=1}^r$ and frequencies $\{\omega_k\}_{k=1}^r$, in the absence of noise, i.e., if $w_l = 0$ for all $l$, the rank of $\f{H}$ is $r$ if $\min(d_1, d_2) \geq r$ (see Theorem 3.2 of Ref.~\cite{Gillard:2023hankel}). This motivates the choice of $\min(d_1, d_2) > r$ to make $\f{H}$ rank-deficient in the noiseless case. As the number of sinusoids $n$, and thus $r$, will be relatively negligible compared to the time series size $L$,  we choose in this work the dimensions $d_1,d_2$ such that the matrix is as close to square as possible and set $d_1 = \lceil L/2 \rceil$. 

Once the dimensions for the embedding are set, there is a one-to-one correspondence between a time series $\f{h}$ and its Hankel matrix $\f{H}$. It is convenient to define the embedding operator $\mathcal{H}: \C^L \rightarrow \C^{d_1 \times d_2}$ which maps $\f{h}$ to $\f{H}$ as defined in Eq.~\eqref{eq:Hankel:def}. The range of $\mathcal{H}$ defines the subspace of Hankel matrices $\mathbb{H}^{d_1 \times d_2}:= \{ \f{X}\in \C^{d_1 \times d_2}: \f{X} = \mathcal{H}(\f{h}), \f{h} \in \C^L\}$. In order to find a denoised version of the time series $\f{h}$, we can do so by embedding the series into a Hankel matrix and finding a low-rank approximation that satisfies the Hankel subspace constraint. Formally, we are thus interested in solving, for a given target rank $r$, the problem
\begin{equation}\label{TargetRankSLRA}
\min_{\f{H} \in \C^{d_1 \times d_2}} \norm{\f{H}-\mathcal{H}(\f{h})}_F, \quad \textrm{s.t.} \ \rank(\f{H}) \leq r, \f{H} \in \mathbb{H}^{d_1 \times d_2},
\end{equation}
where $\norm{\f{H}-\mathcal{H}(\f{h})}_F = \sqrt{\sum_{i,j} |\f{H}_{ij} - \mathcal{H}(\f{h})_{ij}|^2}$ is the Frobenius norm, which is a special case of the \emph{structured low-rank approximation (SLRA)} problem~\cite{Chu:2003slra,Markovsky:2008structured} of approximating the matrix $ \mathcal{H}(\f{h})$ by the closest matrix that is simultaneously of rank limited by $r$ and lies in a specific linear subspace (with the subspace being defined by $\mathbb{H}^{d_1 \times d_2}$). 

In practice, the \textit{target rank} $r$ is chosen based on an estimate of the true number of monochromatic signals composing the time series $\f{h}$ with elements \eqref{eq:timeseries:model}. Unlike an unstructured low-rank approximation problem, which can be solved by a partial singular value decomposition (SVD), direct analytic solutions of \eqref{TargetRankSLRA} are not available~\cite{Markovsky:2008structured}. Apart from choosing a suitable numerical algorithm to solve \eqref{TargetRankSLRA}, which is the focus of the next subsections, the correct choice of the target rank $r$ for a given signal is another relevant problem, which will be further discussed in Section \ref{sec:results}.

Conceptually, the elimination of smaller singular values in the structured low-rank reconstruction corresponds to the elimination or suppression of underlying noise components $w_l$ of the time series \eqref{eq:timeseries:model}  which contribute little to the variance of the series itself.

\subsection{The ESPRIT Algorithm} \label{subsec:esprit}
In signal and array processing, a well-known method to estimate signal parameters such as frequencies and amplitudes in \eqref{eq:timeseries:model} from noisy samples is the ESPRIT (\emph{Estimation of Signal Parameters via Rotational Invariance Techniques}) algorithm~\cite{RoyKailath_ESPRIT:1989}, particularly in the context of array processing. It has the advantage of being computationally efficient as a non-iterative method, and it is one of the preferred methods for direction of arrival estimation~\cite{Qian-2014Computationally,Yang-2018Sparse,Jung-2021Scalable} or time-of-flight estimation~\cite{Othmani-2024ToF}.

The ESPRIT algorithm, outlined in Algorithm~\ref{alg:esprit}, estimates the frequencies $\{\omega_k\}_{k=1}^r$ in what can be considered a two-step projection method onto the two constraint sets of \eqref{TargetRankSLRA}: Step 1 implicitly projects the Hankelized sample vector $\mathcal{H}(\f{h})$ onto the set of matrices with rank at most $r$, and steps 2-5 use the rotational invariance property of the Hankel matrix space $\mathbb{H}^{d_1 \times d_2}$ to extract the parameters $\{\omega_k\}_{k=1}^r$, $\{a_k\}_{k=1}^r$ that can be interpreted as frequency and amplitude vectors. 

Step 5, which estimates the amplitudes, can be motivated by the matrix form
$$
\begin{aligned}
\mathbf{h}
&= \mathbf{B}\mathbf{a} + \mathbf{w},
\end{aligned}
$$
of Eq.~\eqref{eq:timeseries:model}, where $\mathbf{B}=[\mathbf{b}(\omega_1)\cdots\mathbf{b}(\omega_r)]$ is the Vandermonde matrix~\cite{Golub_VanLoan} consisting of columns $\mathbf{b}(\omega_k)=(e^{j\omega_k},\ldots, e^{j\omega_k L})^T$, and where $\mathbf{h}=(h_1,h_{2},\ldots, h_{L})^T$, $\mathbf{a}=(a_1,\ldots,a_r)^T$ and $\mathbf{w}=(w_1,w_{2},\ldots, w_{L})^T$ are vectors encoding the time series, the amplitudes and the noise values, respectively.

\begin{algorithm}[H]
\caption{ESPRIT Algorithm}
\label{alg:esprit}
 \textbf{Input:}
   Samples $\{h_l\}_{l=1}^L$, target rank $r$. \\
 \textbf{Output:} Frequencies $\{\omega_k\}_{k=1}^r$, amplitudes $\{a_k\}_{k=1}^r$.
\begin{algorithmic}[1]
\State  Compute the truncated singular value decomposition $\operatorname{SVD}(\mathcal{H}(\f{h}),r) = \f{U} \Sigma \f{V}^\dagger$ with $\f{U} \in \C^{d_1 \times r}$, $\f{V} \in \C^{d_2 \times r}$ of order $r$ of Hankelized sample vector $
  \mathcal{H}(\f{h}) \in \C^{d_1 \times d_2}$ with $\f{h} = \begin{pmatrix} h_l \end{pmatrix}_{l=1}^L \in \C^{L}$.
\State With $\f{U} %
  \in \C^{d_1 \times r}$ being the matrix of leading left singular vectors of $\mathcal{H}(\f{h})$, we set
  \begin{enumerate}
    \item $\f{U}_1 \in \C^{d_1-1 \times r}$ 
    as matrix of first $(d_1-1)$ rows of $\f{U}$,
    \item $\f{U}_2 \in \C^{d_1-1 \times r}$ as matrix of last $(d_1-1)$ rows of $\f{U}$.
  \end{enumerate}
\State Solve $\Phi = (\f{U}_1^\dagger \f{U}_1)^{-1} \f{U}_1^\dagger \f{U}_2 \in \C^{r \times r}$.
\State Compute eigenvalues $\{\lambda_k\}_{k=1}^r$ of $\Phi$ and set $\omega_k = -j\ln(\lambda_k) \in \C$.
\State Form Vandermonde $\mathbf{B}$ from $\{\omega_k\}_{k=1}^r$ and solve $\mathbf{h} = \mathbf{B} \mathbf{a}$ for $\mathbf{a} = (a_k)_{k=1}^r$.
\State \textbf{return} $\{\omega_k\}_{k=1}^r$, $\{a_k\}_{k=1}^r$.
\end{algorithmic}
\end{algorithm}
While ESPRIT remains a state-of-the-art algorithm for spectral estimation problems for signals of the type \eqref{eq:timeseries:model} in the presence of noise with a well-established theory~\cite{Li:2020esprit,Li:2022stability}, it is an inherently non-iterative algorithm that might not lead to optimal solutions of the underlying SLRA problem \eqref{TargetRankSLRA}.

\subsection{Cadzow Iterations\label{subsec:cadzow}}
An alternative approach for solving the structured low-rank approximation problem \eqref{TargetRankSLRA} is due to Cadzow~\cite{Cadzow-1988,Wang-2021Fast} and has been popularized for solving seismic signal reconstruction~\cite{Trickett-2008,Oropeza-2011Simultaneous} and time series denoising problems~\cite{Golyandina-2005SSA,Gillard:2023hankel}. Unlike ESPRIT, Cadzow's approach (see Algorithm~\ref{alg:cadzow}) is of an iterative nature, and alternately projects onto the space of rank $r$ matrices and onto the subspace of Hankel matrices, respectively, starting from the Hankelized sample vector $\mathcal{H}(\f{h}) \in \mathbb{H}^{d_1 \times d_2}$.

\begin{algorithm}[H]
\caption{Cadzow Iterations}
\label{alg:cadzow}
\textbf{Input:} Samples $\{h_l\}_{l=1}^L$, target rank $r$, maximal number of iterations $T$, convergence threshold $\eta > 0$. \\
\textbf{Output:} Denoised sample vector $\widehat{\f{h}} \in \C^{L}.$
\begin{algorithmic}[1]
\State Set $\f{H}^{(0)} = \mathcal{H}(\f{h})$ for $\f{h} = \begin{pmatrix} h_l \end{pmatrix}_{l=1}^L \in \C^{L}$.
 \For{$t = 1$ to $T$}
    \State Compute the truncated singular value decomposition $\widetilde{\f{H}}^{(t)}:= \operatorname{SVD}(\f{H}^{(t-1)},r) = \f{U} \Sigma \f{V}^\dagger$ with $\f{U} \in \C^{d_1 \times r}$, $\f{V} \in \C^{d_2 \times r}$ of order $r$.
    \State Project $\widetilde{\f{H}}^{(t)}$ to closest Hankel matrix via \emph{antidiagonal averaging} $\f{H}^{(t)}:= \mathcal{P}_{\mathbb{H}^{d_1 \times d_2}}(\widetilde{\f{H}}^{(t)})$, cf. \eqref{eq: AD Averaging}. %
    \State \textbf{if} $\|\f{H}^{(t)} - \f{H}^{(t-1)}\|_F < \eta$ \textbf{then break}
 \EndFor
  \State \textbf{return} $\widehat{\f{h}}:= \y{H}^*(\f{H}^{(t)})$.
\end{algorithmic}
\end{algorithm}
Computationally, step 4 of Algorithm~\ref{alg:cadzow}, which implements the Euclidean projection
\[
\begin{split}
\f{H}^{(t)} &= \mathcal{P}_{\mathbb{H}^{d_1 \times d_2}}(\widetilde{\f{H}}^{(t)}) := \argmin_{\f{H} \in \mathbb{H}^{d_1 \times d_2}} \norm{\f{H}-\widetilde{\f{H}}^{(t)}}_F \\
&= \mathcal{H} (\mathcal{H}^* \mathcal{H})^{-1} \mathcal{H}^* (\widetilde{\f{H}}^{(t)}),
\end{split}
\]
can be implemented as follows. Defining the $k$-th \emph{antidiagonal} of a $(d_1 \times d_2)$ matrix as its entries at the indices $D_k = \{(l_1,l_2): l_1+l_2-1 = k \}$, i.e., the matrix entries whose row index $\ell_1$ and column index $\ell_2$ sum up to $k+1$, $\f{H}^{(t)}$ can be computed by \emph{antidiagonal averaging} of $\widetilde{\f{H}}^{(t)}$: for each antidiagonal $k = 1, \ldots, d_1+d_2-1$, compute
\[
d_k = \frac{1}{|D_{k}|} \sum_{(l_1,l_2) \in {D_{k}}} \widetilde{\f{H}}^{(t)}_{l_1 l_2},
\]
and set the $(\ell_1,\ell_2)$-th entry of $\f{H}^{(t)}$ to
\begin{equation} \label{eq: AD Averaging}
\f{H}_{l_1 l_2}^{(t)} = d_{l_1+l_2-1}.%
\end{equation}
Although this as well as standard routines for partial singular value decomposition allow for a simple implementation of Algorithm~\ref{alg:cadzow}, the Hankel structure can further be leveraged  to implement the partial SVD as well as the projection onto Hankel matrices extremely quickly (each iteration can be performed in $\mathcal{O}(r L\log{L})$ time~\cite{Gillard:2023hankel}): see Sec.~\ref{sec:discussion} for further discussion of implementation details. However, a drawback of Cadzow iterations is that, due to the non-convex nature of the set of matrices of rank at most $r$, it is, even in the case where convergence is achieved, generally unclear whether Algorithm~\ref{alg:cadzow} returns a global minimizer of \eqref{TargetRankSLRA} or not.

\subsection{Iteratively Reweighted Least Squares\label{subsec:IRLS}}
In order to find potentially better and more robust solutions of the structured low-rank approximation problem \eqref{TargetRankSLRA}, we also consider an adaptation of the \emph{iteratively reweighted least squares (IRLS)}~\cite{Fornasier11,KM18_iTwist18,KM19,KuemmerleMayrinkVerdun-ICML2021} framework to the denoising of time series. Note that although we have so far designed LISA time series methodologies by modeling the problem within \eqref{TargetRankSLRA}, IRLS can be derived from the related \emph{structured rank minimization} formulation, in which, for given Frobenius error tolerance $\eta>0$, we look for solutions of
\begin{equation}
\min_{\f{H} \in \mathbb{H}^{d_1 \times d_2}} \rank(\f{H}), \quad \textrm{s.t.} \ \norm{\f{H}-\mathcal{H}(\f{h})}_F < \eta.
\label{eq:HankMin}
\end{equation}
A solution of \eqref{eq:HankMin} would provide us the Hankel matrix close to the embedding of the noisy time series, but also corresponding to a sum of a much lower number of monochromatic signals. This would allow one to extract the most important signals from the data matrix. \\
Unfortunately, the non-convex and discontinuous nature of the rank objective within \eqref{eq:HankMin} makes a direct minimization very challenging \cite{Recht2010guaranteed}. This challenge is mitigated by considering a continuously differentiable, but similarly non-convex surrogate rank function with similar minimizers such as the \emph{$\varepsilon$-smoothed log-determinant rank surrogate} defined with respect to the singular values $\sigma_i(\f{H})$ of $\f{H}$ as
\begin{equation} \label{eq:surrogate:definition}
\y{J}_\varepsilon(\f{H}) = \sum_{i=1}^{\min(d_1,d_2)} j_\varepsilon(\sigma_i(\f{H})),
\end{equation}
where $j_\varepsilon: \R_{\geq 0} \to \R$ is such that
\[
j_\varepsilon(\sigma) = \begin{cases}
\log(\sigma), \ \ &\text{if} \ \  \sigma> \varepsilon, \\
\log(\varepsilon) + \frac{1}{2}\left(\frac{\sigma^2}{\varepsilon^2}-1\right), \ \ &\text{if} \ \ 0\leq \sigma \leq \varepsilon.
\end{cases}
\]
This smoothed surrogate function mimics the step-discontinuity of the rank function while relaxing the minimization problem such that it is feasible to solve. Rather than optimizing this function directly, we minimize via iteratively reweighted least squares (IRLS). At each iteration, the algorithm optimizes over a quadratic (second-order) surrogate of 
$\y{J}_\varepsilon(\f{H})$ centered about the Hankelized estimate $\f{H}^{(t)} = \mathcal{H}(\f{h}^{(t)})$ of the time series vector $\f{h}^{(t)}$:
\begin{equation} \label{eq:quad:model}
\begin{split}
Q_{\varepsilon,\f{H}^{(t)}}(\f{H}) 
    &= \y{J}_\varepsilon\!\big(\f{H}^{(t)}\big) 
    + \,\big\langle \nabla \y{J}_\varepsilon\!\big(\f{H}^{(t)}\big),\, \f{H} - \f{H}^{(t)} \big\rangle \\
    &\quad + \tfrac{1}{2}\,\big\langle \f{H} - \f{H}^{(t)},\, 
        W_\varepsilon^{(t)}\!\big(\f{H} - \f{H}^{(t)}\big) \big\rangle,
\end{split}
\end{equation}
where the (linear) weight operator $W_\varepsilon^{(t)}: \mathbb{H}^{d_1 \times d_2} \to \C^{d_1 \times d_2}$ is defined by
\begin{equation}\label{Wop}
\begin{split}
    W_\varepsilon^{(t)}(\f{H}) 
    &\;=%
      \f{U}^{(t)} 
\Sigma_{\varepsilon,d_1}^{(t)-1} \f{U}^{(t)\dagger} \f{H} \f{V}^{(t)} \Sigma_{\varepsilon,d_2}^{(t)-1} \f{V}^{(t)\dagger}\,,
\end{split}
\end{equation}
where $\f{U}^{(t)}$, $\f{V}^{(t)}$ and $\Sigma^{(t)} = \diag(\sigma_i(\f{H}^{(t)})_{i=1}^{\min(d_1, d_2)}) \in \C^{d_1 \times d_2}$ are obtained from the singular value decomposition $\f{H}^{(t)} = \f{U}^{(t)} \Sigma^{(t)} \f{V}^{(t)\dagger}$ of $\f{H}^{(t)}$, and $\Sigma_{\varepsilon,d}^{(t)} =\max(\Sigma^{(t)}, \varepsilon \mathbf{I}) = \diag(\max(\sigma_i(\f{H}^{(t)}), \varepsilon)_{i=1}^d)$. 

The structure of the weight operator $W_\varepsilon^{(t)}$ is related to the Hessian $\nabla^2 \y{J}_{\varepsilon}(\f{H}^{(t)})$ of the smoothed surrogate objective $\y{J}_\varepsilon(\cdot)$ such that $W_\varepsilon^{(t)} \succeq \nabla^2 \y{J}_{\varepsilon}(\f{H}^{(t)})$ where $\succeq$ denotes that $\ip{\f{H}}{W_\varepsilon^{(t)}(\f{H})} \geq \ip{\f{H}}{\nabla^2 \y{J}_{\varepsilon}(\f{H}^{(t)})(\f{H})}$ for all $\f{H} \in \C^{d_1 \times d_2}$, while simultaneously ensuring that the associated quadratic model function \eqref{eq:quad:model} is a tight upper bound for $\y{J}_\varepsilon(\cdot)$ for any $\f{H}^{(t)}$ \cite{KuemmerleMayrinkVerdun-ICML2021}. 

This derivation results in a tractable iterative scheme by relaxing \eqref{eq:HankMin} into 
\[
\min_{\f{\hat{h}} \in \C^{L}} Q_{\varepsilon,\y{H}(\f{h}^{(t)})}(\y{H}(\f{\hat{h}})) \ \textrm{s.t.} \ \norm{\y{H}(\f{\hat{h}})-\mathcal{H}(\f{h})}_F < \eta.
\]
Although this is a convex quadratically constrained quadratic program (QCQP), it remains challenging to solve; we therefore consider the unconstrained variant of the problem 
\begin{equation}\label{IRLSFun}
\min_{\f{\hat{h}} \in \C^L} \lambda Q_{\varepsilon,\y{H}(\f{h}^{(t)})}(\y{H}(\f{\hat{h}}))+\frac{1}{2}\norm{\f{\hat{h}}-\f{h}}_2^2,
\end{equation}
parametrized by a positive regularization parameter $\lambda > 0$.

The target rank $r$, which was used in Algorithm~\ref{alg:esprit} and Algorithm~\ref{alg:cadzow} as a hard constraint, can now be used to guide the rank selection \emph{smoothly} by being part of the update rule of the smoothing parameter $\varepsilon$ (cf. step 6 of Algorithm~\ref{alg:HankIRLS}) in \eqref{eq:surrogate:definition}, recognizing the fact that gradually reducing the smoothing improves the solution quality in IRLS algorithms with non-convex objectives \cite{Wipf10,KuemmerleMayrinkVerdun-ICML2021}.

These considerations allow us to formulate an IRLS algorithm for time series denoising, detailed in Algorithm~\ref{alg:HankIRLS}, observing that the minimization~\eqref{IRLSFun} is equivalent to finding the unique minimizer of the quadratic form
\begin{equation} \label{eq:IRLS}
    \f{h}^{(t+1)} = \operatorname{arg}\min_{\f{\hat{h}} \in \C^L} \ \frac{\lambda}{2}\ip{\f{\hat{h}}}{\mathcal{H}^*W_\varepsilon^{(t)}\mathcal{H}(\f{\hat{h}})} + \frac{1}{2}\norm{\f{\hat{h}}-\f{h}}_2^2
\end{equation}
among candidate time series vectors $\f{\hat{h}} \in \C^L$. Compared to \eqref{eq:HankMin}, we use the adjoint of the Hankel operator to obtain a problem with an $L$-dimensional vector variable rather than a matrix-valued one, which unlocks certain computational efficiencies. %

\begin{algorithm}[H]
    \caption{IRLS for Time Series Denoising}\label{alg:IRLS}
    \textbf{Input:} 
    Samples $\{h_l\}_{l=1}^L$, target rank $r$, 
    maximal number of iterations $T$, regularization $\lambda$, relative tolerance $\tau$ \\
    \textbf{Output:} Denoised sample vector $\widehat{\f{h}} \in \C^{L}$.
    \begin{algorithmic}[1]
        \State Set $\f{h}^{(0)} := \f{h} := \begin{pmatrix} h_l \end{pmatrix}_{l=1}^L \in \C^{L}$ and $\varepsilon = +\infty$.
        \State Initialize $W_{\varepsilon}^{(0)}$ as identity operator. %
        \For{$t=0$ to $T-1$}
            \State Compute solution $\f{h}^{(t+1)} \in \C^L$ of least-squares problem \eqref{eq:IRLS}.%
            \State Compute $\f{U},\Sigma,\f{V} \;=\; \operatorname{ParSVD}(\mathcal{H}(\f{h}^{(t+1)}),\varepsilon,r+1)$ to define weight operator $W_{\varepsilon}^{(t+1)}(\cdot)$ via \eqref{Wop}.
        
            \State Set $\varepsilon \;:=\; \min\big(\Sigma_{r+1,r+1},\, \varepsilon\big)$
             \State \textbf{if} $\frac{\|\f{h}^{(t+1)} - \f{h}^{(t)}\|_2}{\|\f{h}^{(t)}\|_2} < \tau $ \textbf{then break}
        \EndFor
        \State \textbf{Return:} $\widehat{\f{h}}: = \f{h}^{(t+1)}$
      \end{algorithmic}
      \label{alg:HankIRLS}
\end{algorithm}

To further refine the algorithm's convergence toward a low-rank solution, we implement an adaptive schedule for the regularization parameter $\lambda$. This parameter is responsible for the trade-off between the target rank (the first term on the r.h.s. of~\eqref{eq:IRLS}) and the data fidelity (the second term). A small $\lambda$ allows the model to fit noise fluctuations (overfitting), resulting in a high-rank matrix, whereas a large $\lambda$ enforces the rank constraint too aggressively, potentially leading to a poor approximation of the signal (underfitting). As a result, a fixed $\lambda$ proves quite restrictive for driving our algorithm towards a meaningful solution.

In this refinement step, we initialize~$\lambda$ with a relatively small value ($\lambda_0=0.1$) to prioritize fitting the data in the early iterations. We then monitor the leakage into higher singular values (which should be negligible for a low-rank approximation) by calculating the spectral tail ratio,
\begin{equation}
\beta_{\text{spec}} = \frac{\sqrt{\sum_{i=1}^{r+1} \sigma_i^2}}{\|\mathcal{H}(\f{h}^{(t)})\|_F}\,.
\end{equation}
If the solution meets the tolerance condition in Step~7 but $\beta_{\text{spec}}$ remains below a threshold close to unity\footnote{For any matrix, the Frobenius norm is equal to the Euclidean norm (sum in quadrature) of its singular values. Therefore, if $\sigma_i \approx 0$ for all $i > r$, this ratio must be close to unity.}, we increase $\lambda$ as $\lambda_{k+1} = 1.2\lambda_{k}$ and continue the iterations. This strategy progressively enforces the low-rank constraint more aggressively, slowly eliminating the residual noise. More sophisticated adaptive schedulers are left to future work.

In step 5 of Algorithm~\ref{alg:IRLS}, $\operatorname{ParSVD}(\mathcal{H}(\f{h}^{(t+1)}),\varepsilon,r+1)$ denotes computing a partial (truncated) singular value decomposition of $\mathcal{H}(\f{h}^{(t+1)})$ such that at least $\min(r+1, \min(d_1,d_2))$ singular value triplets are calculated. While not obvious, the spectral information of $O(r)$ singular triplets calculated in this step is sufficient to define the updated weight operator $W_{\varepsilon}^{(t+1)}$ and the updated weighted least squares problem \eqref{eq:IRLS} (see, e.g., \cite{ghosh2025qr}).
 We postpone a discussion of the computational complexity and our implementation to Section \ref{sec:discussion}.

 Compared to Cadzow's approach or ESPRIT, the IRLS method of Algorithm~\ref{alg:HankIRLS} has the potential for finding better solutions of the underlying structured low-rank approximation problem through its iterative nature involving the optimization of non-convex, but continuous rank surrogates with decreasing smoothing parameter.

Whenever interested in identifying amplitudes and frequencies relevant for the time series at hand, for both Cadzow and IRLS, it is possible to postprocess the resulting \emph{denoised} time series $\widehat{\f{h}}$ with steps 2-5 of Algorithm~\ref{alg:esprit} to obtain these. We follow that approach when reporting frequency estimates in experimental results. 
\begin{figure*}[htbp!]
\centering
\includegraphics[width=0.45\textwidth]{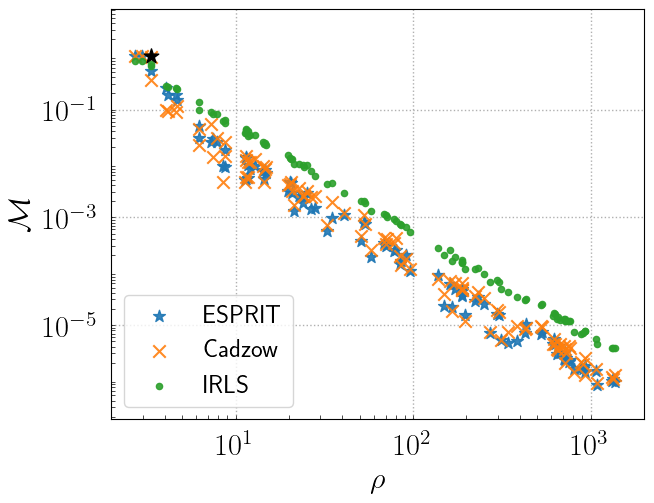}
\includegraphics[width=0.99\textwidth]{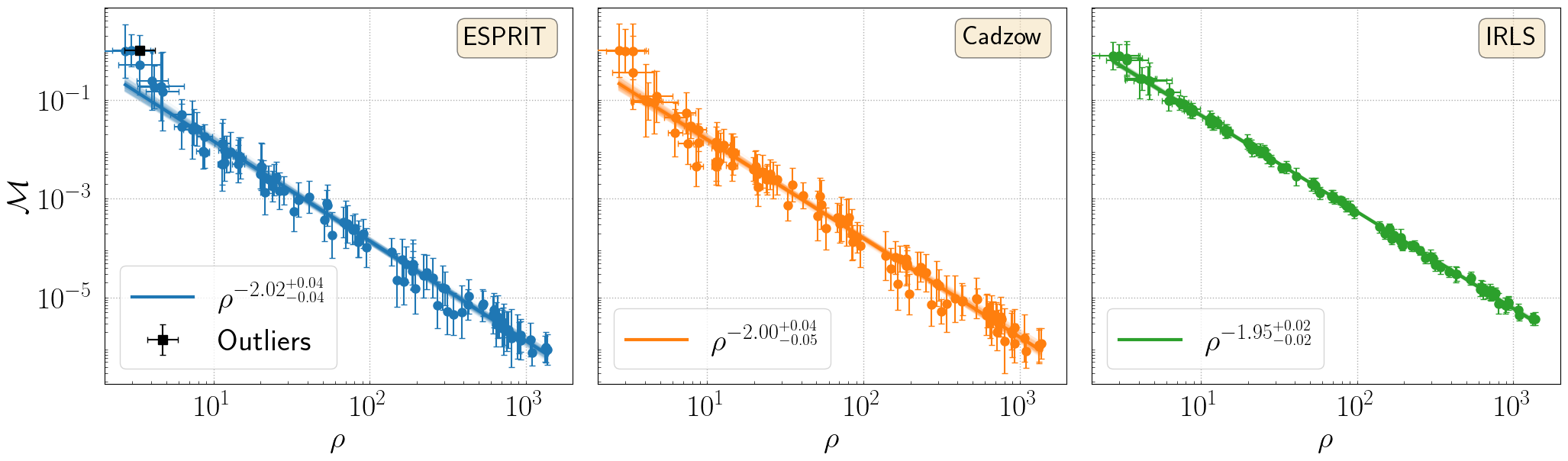}
    \caption{Mismatch~$\mathcal{M}$ vs. SNR~$\rho$ for the single monochromatic signal experiment. Top: Combined results for ESPRIT (stars), Cadzow (crosses), and IRLS (dots). Bottom: Individual results for each algorithm, where data points are the medians over 50 noise realizations and error bars span the 16th-84th percentiles. The solid lines represent the best-fit power law obtained from a Markov chain Monte Carlo (MCMC) procedure (see Eq.~(17) of~\cite{Hogg:2010yz}), with faint lines indicating samples from the posterior. Outliers identified by this procedure (posterior probability $>0.5$) are marked in black using the respective algorithm's symbol.}
\label{fig:single_mismatch}
\end{figure*}

\section{Results\label{sec:results}}

\begin{figure}[htbp!]
\centering
\includegraphics[width=0.45\textwidth]{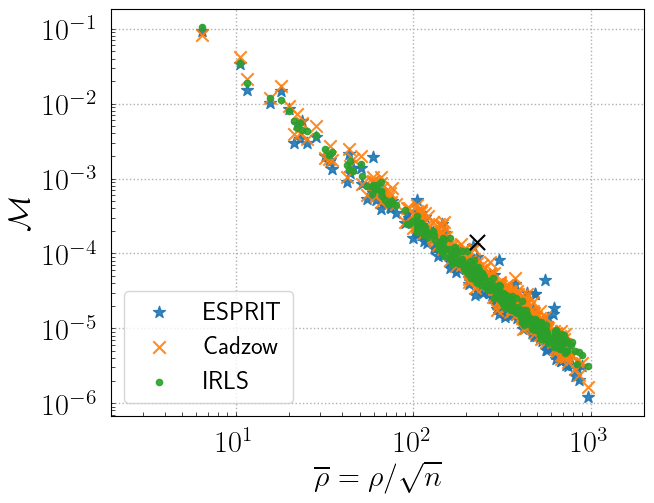}
    \caption{Mismatch $\mathcal{M}$ vs. the normalized SNR $\overline{\rho} = \rho/\sqrt{n}$ for the multiple-signal experiment with a known number of signals. This plot combines results for superpositions of $n=3$, $5$ and $7$ signals, denoised with ESPRIT (blue stars), Cadzow (orange crosses), and IRLS (green dots). The fact that all data points follow the universal trend $\mathcal{M} \propto \overline{\rho}\,{}^{-2}$ confirms that the denoising performance for all three algorithms scales as expected. Qualitatively, the scatter is smallest for IRLS and largest for ESPRIT. This scatter, as well as the outliers, typically corresponds to superpositions where some signals are significantly weaker than the others.}
\label{fig:multi_mismatch}
\end{figure}

\begin{figure}[htbp!]
\centering
\includegraphics[width=0.95\columnwidth]{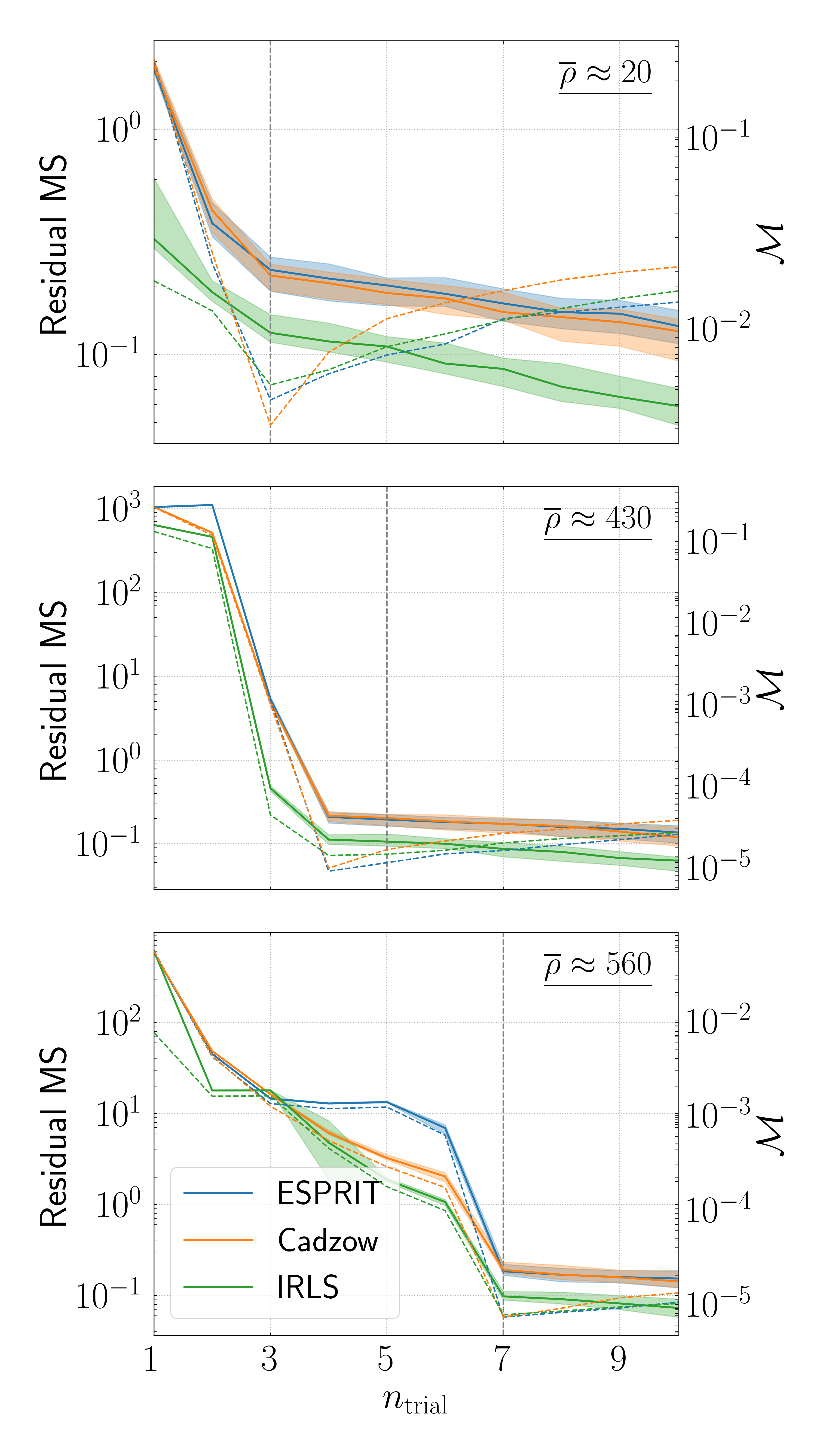}
    \caption{Mismatch $\mathcal{M}$ (right $y$-axis, dashed lines) and the mean square (MS) of the residual (left $y$-axis, solid lines) as a function of the trial number of components~$n_{\text{trial}}$, for experiments with an unknown number of signals. The three panels correspond to experiments with a true number of signals $n_{\text{true}} = 3, 5,$ and $7$ (top to bottom). For all algorithms, the mismatch reaches a minimum when $n_{\text{trial}}$ approximately matches the ground truth (vertical dashed line). The residual MS exhibits an ``elbow'' at this point before continuing to decrease, which is a signature of overfitting. Interestingly, in the middle panel, the minima and elbows correctly indicate only four components, as one of the five signals in this particular realization was too weak for the algorithms to resolve. The normalized SNRs of randomly generated superpositions are indicated in each panel.}
\label{fig:sweep_mismatch}
\end{figure}

In this section, we present the results of the numerical experiments described in Sections~\ref{sec:intro} and~\ref{sec:setup}. For the one- and multiple-signal scenarios, we use the mismatch $\mathcal{M}$ \eqref{eq:mismatch} as a performance metric and plot it as a function of the normalized SNR $\overline{\rho}$, as defined in~\eqref{eq:normalized_snr}. In the case of one signal, it simply coincides with the SNR of the signal~$\rho$. For the frequency separation experiment, we evaluate the algorithms using the relative root-mean-square (r.m.s.) error in the reconstructed frequencies, $\sigma_f = \sqrt{(\Delta f_1/f_1)^2 + (\Delta f_2/f_2)^2}$, as well as the scaled mismatch $\mathcal{M}\overline{\rho}^2$. 

Finally, since in the QNM experiment our goal is to test the denoising techniques as a proof of concept, we simply plot the extracted complex frequencies $M_f \omega$ across the start times (between $10M$ and $40M$ after the peak). At the relatively high median SNR considered here, we expect these frequencies to cluster around the theoretical predictions. This clustering should persist even as we vary the start time over a wide range.

\subsection{One Monochromatic Signal\label{subsec:single}}

The results for the one-signal experiment, which serves as a consistency check for our methods, are shown in Fig.~\ref{fig:single_mismatch}. The top panel plots the mismatch $\mathcal{M}$ as a function of SNR $\rho$ for all three algorithms, where each point is the median over $50$~noise realizations. The SNR range between $\approx 3$ and $\sim 10^3$ arises from sampling the signal amplitude from a log-uniform distribution while maintaining a constant noise PSD, $S_{\rm n}=1$. For an arbitrary PSD, the data can be whitened, which involves dividing by the amplitude spectral density $\sqrt{S_{\rm n}(f)}$ in the Fourier domain. For monochromatic signals, this results in a scaling factor for the amplitude at the frequency of the signal, $a\to a/\sqrt{S_{\rm n}(f_0)}$.

Individual results for each algorithm are presented in the bottom panels of Fig.~\ref{fig:single_mismatch}. The data points are the same as in the top panel but now shown with error bars spanning the 16th-84th percentiles over the noise realizations, as well as a linear fit in log-log space. The fit is performed using an MCMC procedure that models the data as a mixture of inliers and outliers, following Eq.~(17) of~\cite{Hogg:2010yz}. A point is classified as an outlier if its posterior probability exceeds a threshold of 0.5. These are marked in black using the respective algorithm's marker. 

All methods clearly exhibit the expected inverse-square relationship, $\mathcal{M} \propto \rho^{-2}$, which is consistent with the theoretical prediction from Fisher matrix analysis, Eq.~\eqref{eq:fisher_prediction}. Specifically, the best-fit power-law exponents are $-2.02_{-0.04}^{+0.04}$ for ESPRIT, $-2.00_{-0.05}^{+0.04}$ for Cadzow, and $-1.95_{-0.02}^{+0.02}$ for IRLS. This confirms that the denoising performance is close to optimal for all three algorithms. 

However, the algorithms also exhibit notable differences at low SNR. On one hand, ESPRIT and Cadzow demonstrate deviations from the inverse-square dependence below $\rho\approx 5$ and $\rho\approx 7$, respectively, and we occasionally observe one or two outliers below those thresholds. This behavior is expected, because the Fisher matrix analysis is known to be valid only in the limit of high SNRs~\cite{Vallisneri:2007ev}. The low-SNR thresholds we find are also consistent with the cutoff values typically assumed when estimating observable populations of DWDs for LISA (e.g.,~\cite{Karnesis:2021tsh,Korol:2021pun}). For IRLS, on the other hand, the inverse-square relationship extends smoothly to the lower end of the SNR range. The absence of a low-SNR breakdown is due to the systematic offset present in the IRLS linear fit. In the regime of low SNR, the absolute mismatch values are comparable for all three algorithms. As discussed in Section~\ref{sec:discussion}, this offset (which is statistically significant) is a direct consequence of the algorithmic regularization. The regularization parameter $\lambda$ in IRLS appears to favor the low-rank constraint, thus underfitting the data.

\begin{table}[h!]
    \centering
    \caption{Summary of the best-fit power-law exponents, $\alpha$, for the mismatch scaling relationship $\mathcal{M} \propto \overline{\rho}\,{}^{\alpha}$ and different number of signals~$n$. The values are obtained as described in Section~\ref{subsec:multi}. In the case of one monochromatic signal, the power-law exponents are the same as reported in Fig.~\ref{fig:single_mismatch}.\label{tab:fit_exponents}}
    \begin{tabular}{l|ccc}
        \hline\hline
         \\[-0.8em]
         & ESPRIT & Cadzow & IRLS \\[1ex] \hline
         \\[-0.8em]
        $n=1$ & $-2.02_{-0.04}^{+0.04}$  & $-2.00_{-0.05}^{+0.04}$ & $-1.95_{-0.02}^{+0.02}$ \\[1ex] 
        $n=3$ & $-2.08_{-0.04}^{+0.04}$  & $-2.09_{-0.03}^{+0.04}$ & $-1.97_{-0.02}^{+0.02}$ \\[1ex]
        $n=5$ & $-2.05_{-0.04}^{+0.04}$  & $-2.04_{-0.04}^{+0.04}$ & $-2.01_{-0.03}^{+0.03}$ \\[1ex]
        $n=7$ & $-1.99_{-0.04}^{+0.04}$  & $-2.03_{-0.04}^{+0.03}$ & $-2.00_{-0.03}^{+0.03}$ \\[1ex]
        \hline\hline
    \end{tabular}
\end{table}

\subsection{Multiple Signals\label{subsec:multi}}

In this experiment, we consider two cases. We first assume that the number of signals~$n$ is known in advance. The results for this case are presented in Fig.~\ref{fig:multi_mismatch}, which shows a summary plot for superpositions of $n = 3, 5$, and $7$ signals. For each algorithm, we plot the mismatch $\mathcal{M}$ as a function of the normalized SNR, $\overline{\rho} = \rho/\sqrt{n}$, combining data points for all superpositions. As expected from Eq.~\eqref{eq:fisher_prediction}, the data points for each algorithm follow a universal trend, which is again consistent with $\mathcal{M} \propto \overline{\rho}\,{}^{-2}$. We provide the best-fit power-law exponents for the individual values of~$n$ in Table~\ref{tab:fit_exponents}. While all algorithms perform well, the scatter in the IRLS data points (green dots) is visibly smaller than that for Cadzow (orange crosses), which in turn is smaller than for ESPRIT (blue stars). We observe that ESPRIT produces more outliers on average, particularly at higher SNR. Although this seems counterintuitive, we find that it typically occurs when one or more signals in the superposition have significantly lower amplitudes than the others, causing ESPRIT to fail in resolving the weaker components. The advantage of Cadzow and IRLS in this experiment may be due to their iterative nature.

We next consider the more realistic scenario where the number of signals~$n_{\text{true}}$ is unknown. To estimate this number, we vary the target number of components in the denoising algorithm, $n_{\text{trial}}$, and evaluate the performance using the mismatch and the mean square (MS) of the residual $(h - \hat{h})$. In real observations, the ground-truth waveform is unavailable, and the residual MS serves as a proxy for the PSD---that is, for a correctly implemented denoising procedure with the correct number of components, the residual should consist only of noise. Assuming ergodicity, where the mean over noise realizations equals the mean over data points in a single realization, one can use the residual MS as an estimate for the noise PSD.

The results for this second case are shown in Fig.~\ref{fig:sweep_mismatch}, where the panels from top to bottom correspond to $n_{\text{true}}=3, 5$, and $7$ (indicated by the vertical dashed line in each plot). The signals for a given superposition were generated randomly, but the same set was used for all three algorithms. As expected, the mismatch (right $y$-axis; dashed lines) exhibits a minimum when the trial number of components matches the ground truth. In contrast, the residual MS (left $y$-axis; solid lines) continues to decrease as $n_{\text{trial}}$ increases. The rate of decrease, however, noticeably changes at $n_{\text{trial}} \approx n_{\text{true}}$, forming an ``elbow'' in the curve. For $n_{\text{trial}} > n_{\text{true}}$, the algorithms begin to overfit the time series: they essentially fit noise fluctuations in addition to the true signal components and continue to reduce the residual MS. This suggests that the location of the ``elbow'' in the residual MS plot can serve as a heuristic for estimating the number of signals in real data.

Note also that, in the middle panel, both the mismatch and residual MS indicate the presence of $4$ components instead of the ground truth $n_{\text{true}}=5$. In this particular realization, one of the five signals was generated with a much smaller amplitude than the other four, and the algorithms were unable to resolve it from the noise.

\subsection{Frequency Separation\label{subsec:separation}}

\begin{figure*}[htbp!]
\centering
\includegraphics[width=0.9\textwidth]{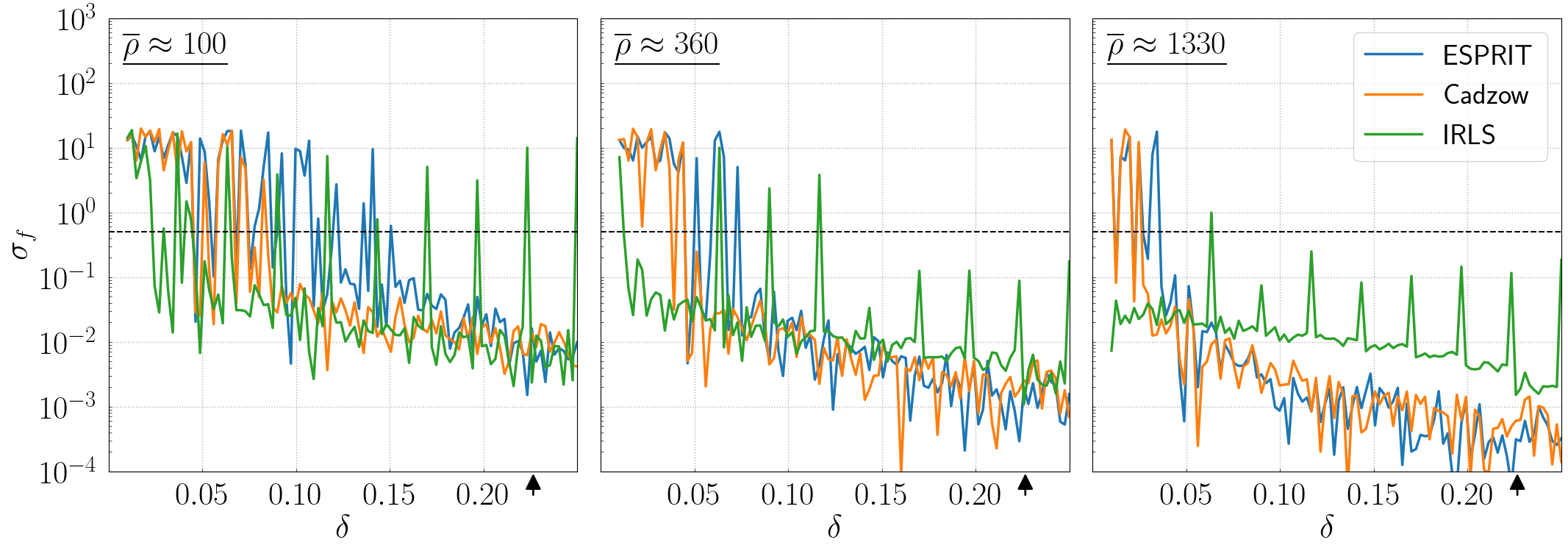}
\includegraphics[width=0.9\textwidth]{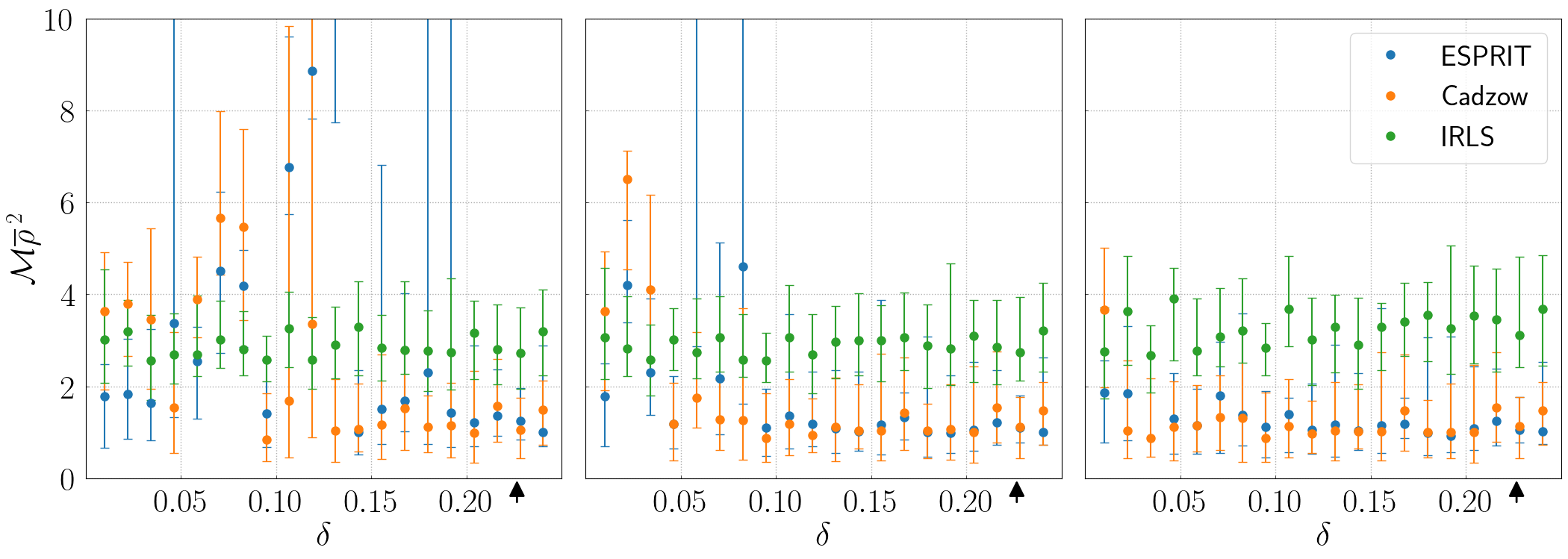}
\caption{
    Denoising performance as a function of the relative frequency separation $\delta = |f_2-f_1|/f_1$. 
    Top: The median relative r.m.s. error in frequency reconstruction~$\sigma_f$ for ESPRIT (blue), Cadzow (orange), and IRLS (green) algorithms across runs with low, moderate, and high SNRs (from left to right). The horizontal dashed line marks a threshold $\sigma_f=0.5$. The approximate SNR indicated in the top left refers to both panels in a column. Bottom: The scaled mismatch $\mathcal{M}\overline{\rho}^2$ for the same set of runs and algorithms. Points represent the median over 50 noise realizations, and error bars span the 16th-84th percentiles. In all panels, the position of the Fourier resolution limit is marked by a black arrow on the horizontal axis.
}
\label{fig:freq_sep}
\end{figure*}

In this experiment, we test how well the algorithms can resolve two monochromatic signals with similar frequencies. As described in Section~\ref{sec:setup}, the signal for this experiment consists of two sinusoids with frequencies $f_1$ and $f_2 = f_1(1 \pm \delta)$, where the relative separation $\delta$ varies between $0.01$ and $0.25$. To control the signal's SNR, we divide the interval between $10a_{\rm min}=2$ and $a_{\rm max}=100$ into three equal parts on a logarithmic scale to sample signals with low, moderate, and high SNRs, respectively. The two signals are assumed to have the same amplitude.

In Fig.~\ref{fig:freq_sep} we show the results of this experiment. As measures of performance, we use the relative r.m.s. error in frequency reconstruction, $\sigma_f = \sqrt{(\Delta f_1/f_1)^2 + (\Delta f_2/f_2)^2}$ (top panel), and the scaled mismatch $\mathcal{M}\overline{\rho}^2$ (bottom panel). The frequency reconstruction error $\sigma_f$ is computed in two steps: (i) we first compute the median of the denoised time series across all noise realizations, and (ii) we apply the ESPRIT algorithm to this median signal to estimate~$f_1$ and~$f_2$.

For an optimally performing algorithm, the scaled mismatch should be constant and of order unity, as follows from Eq.~\eqref{eq:fisher_prediction} and is confirmed by the results of experiments~1 and~2 (see Sections~\ref{subsec:single} and~\ref{subsec:multi}). In general, $\sigma_f$ tracks the scaled mismatch except for very small frequency separations, when the mismatch may remain low while $\sigma_f$ becomes large. This happens because the beat period of the two sinusoids ($T_{\text{beat}} = 1/\Delta f$) becomes much longer than the observation time, thus making the true signal look nearly identical to a single sinusoid. A denoising algorithm may correctly recover a single-sinusoid waveform that has a low mismatch with the true signal, but the subsequent frequency estimation fails to find two distinct frequencies. Thus, a large error (e.g., $\sigma_f \gtrsim 0.5$, indicated by the horizontal dashed line in the top panel) implies a failure to resolve the two components.

The columns in Fig.~\ref{fig:freq_sep} correspond to runs with low, moderate, and high SNRs (from left to right). All of the algorithms appear to demonstrate ``super-resolution'': they are capable of resolving frequency separations $\delta$ below half the size of a Fourier frequency bin $\Delta f<1/(2T)$ (indicated by the black arrow). However, as the SNR decreases, the point where both the mismatch and frequency error begin to degrade shifts to larger values of $\delta$ and closer to the Fourier limit. We leave a detailed study of how the ``super-resolution'' scales with SNR to future work.

Comparing the algorithms, we observe that the iterative Cadzow routine demonstrates a somewhat lower mismatch and frequency reconstruction error than the non-iterative ESPRIT, which is especially apparent at lower SNR. At the same time, the IRLS routine exhibits a higher error compared to the other two algorithms. This is most evident in the top panel, where the IRLS curve for $\sigma_f$ displays sharp peaks that are not present in the ESPRIT and Cadzow results. This, however, does not prevent IRLS from reconstructing frequencies with a relative accuracy better than $30\%$ at higher SNRs. Note also that the higher error for IRLS manifests itself in the systematically higher scaled mismatch in the bottom panel, which is consistent with the offset observed in the one-signal experiment (see Fig.~\ref{fig:single_mismatch}).

\subsection{Quasinormal Modes\label{subsec:quasinorm}}

\begin{figure*}[htbp!]
\centering
\includegraphics[width=0.75\textwidth]{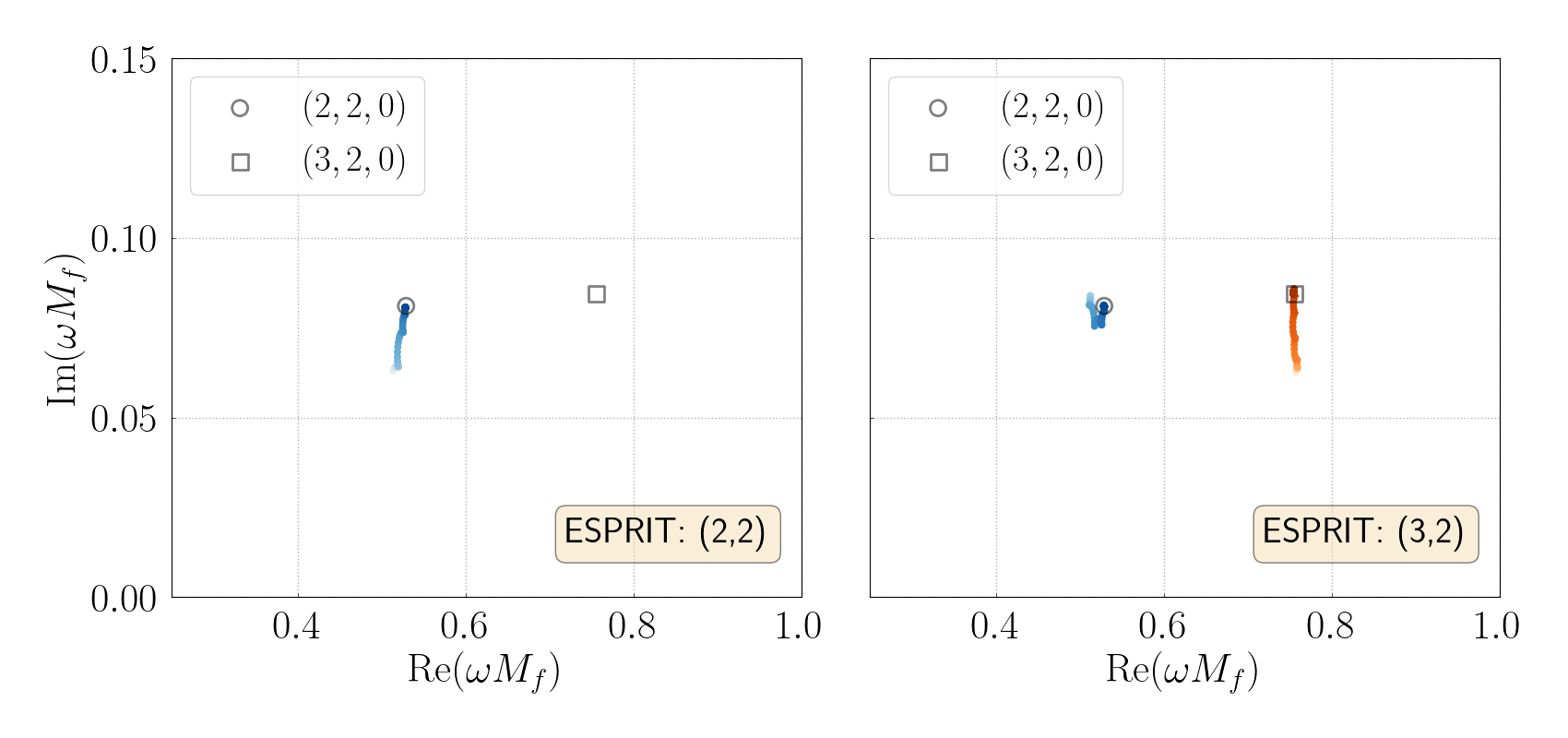}
\includegraphics[width=0.75\textwidth]{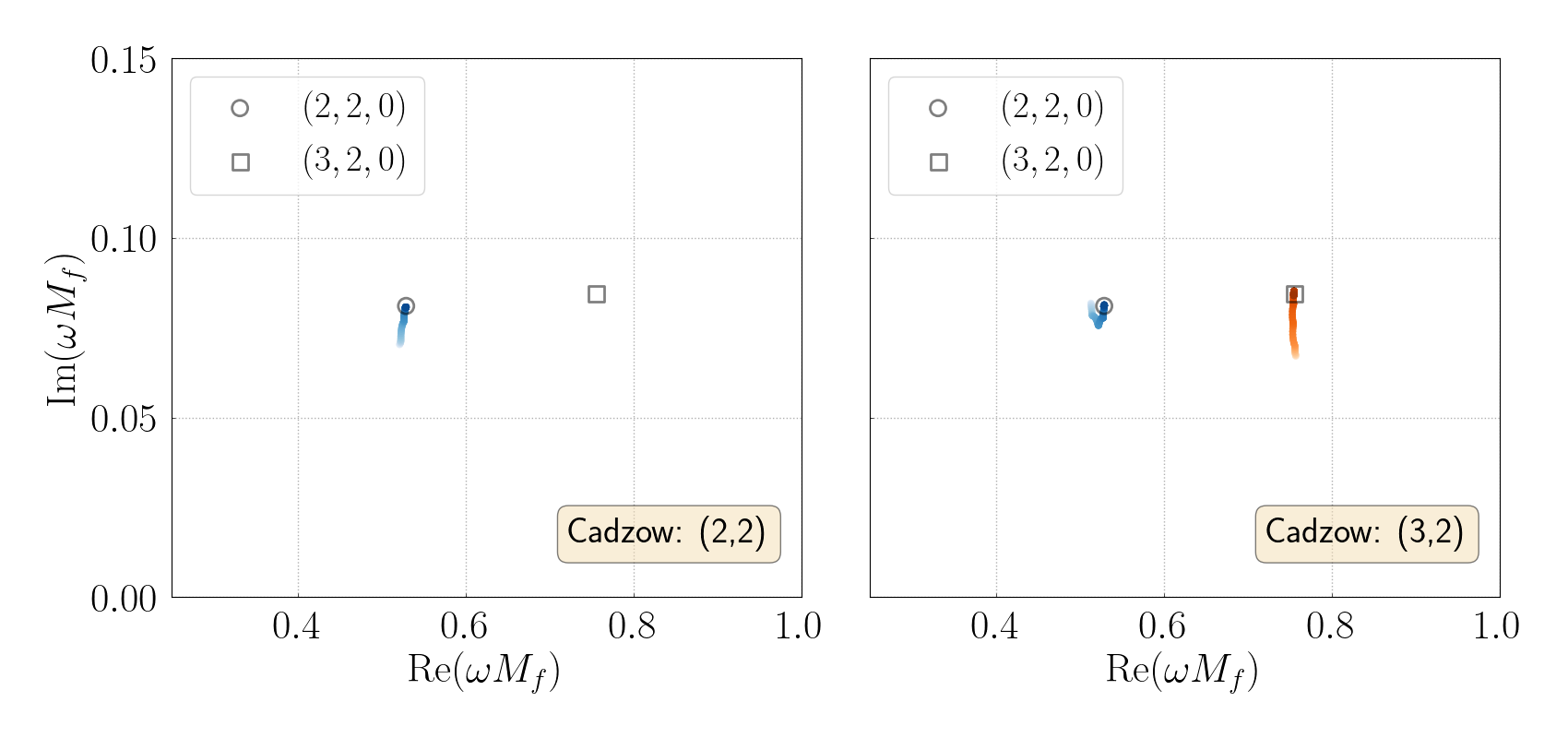}
\caption{
    QNM spectroscopy of the SXS waveform \texttt{SXS:BBH:0305} in the absence of synthetic noise. The left and right columns show results for the spherical harmonic modes $(2,2)$ and $(3,2)$, respectively. The blue and orange tracks represent frequencies extracted by ESPRIT (top) and Cadzow iterations (bottom) across varying start times. The color shading indicates the start time offset, with darker shades corresponding to later start times and shorter ringdown tails. The open markers indicate the theoretical complex QNM frequencies of the fundamental $(\ell, m, n_{\rm tone})=(220)$ mode (circles) and of the $(320)$ mode (squares).
}
\label{fig:qnm_results}
\end{figure*}

In our final experiment, we apply the ESPRIT algorithm to the ringdown phase of the numerical relativity waveform \texttt{SXS:BBH:0305}, which has been extensively used to test ringdown fitting algorithms~\cite{Giesler:2019uxc,Baibhav:2023clw,Cheung:2023vki,Takahashi:2023tkb,Giesler:2024hcr}. We consider two configurations: applying ESPRIT directly, and applying it after the waveform has been preprocessed with Cadzow denoising. In the second case, we set the stopping tolerance to the double floating-point precision ($10^{-16}$) to remove any numerical noise that may be present in the simulation. As detailed in Section~\ref{sec:setup}, the \texttt{SXS:BBH:0305} simulation corresponds to a quasicircular binary black hole system with mass ratio $q=1.22$ and anti-aligned dimensionless spins $\chi_1=0.33$ and $\chi_2=-0.44$, resulting in a final remnant with mass $M_f = 0.9520\,M$ and spin $\chi_f = 0.6921$. We analyze the plus polarization of the $(2,2)$ and $(3,2)$ spherical harmonic modes of the Weyl scalar $\Psi_4$, and in each case we consider the ringdown tail with a start time varying between the peak $t_{\rm peak}$ and $t_{\rm peak}+30M$.

The decomposition of the gravitational radiation into spin-weighted \textit{spherical} harmonics means that a single physical QNM, given in terms of the spin-weighted \textit{spheroidal} harmonics~\cite{Teukolsky:1973ha,Leaver:1985ax}, contributes to multiple spherical modes. This mode mixing is particularly relevant for the $(3,2)$ spherical mode~\cite{Buonanno:2006ui,Berti:2007fi,Kelly:2012nd,Berti:2014fga}. While the $(2,2)$ data mode is overwhelmingly dominated by the fundamental $(\ell, m, n_{\rm tone}) = (220)$ QNM, the $(3,2)$ mode contains a superposition of its ``native'' $(320)$ QNM and the leakage from the much louder $(220)$ mode. This mode mixing is typically visible in the time domain as an amplitude modulation (beating). Accordingly, we set the target number of signals/frequencies to recover as $n=1$ for the $(2,2)$ data mode and $n=2$ for the $(3,2)$ data mode. To compare our results to the ground truth, we calculate the theoretical frequencies and the dominant spherical-spheroidal mixing coefficients using the \texttt{qnm} Python package~\cite{Stein:2019mop}.

Figure~\ref{fig:qnm_results} shows the results of this experiment. The left column corresponds to the dominant $(2,2)$ data mode, while the right column shows the $(3,2)$ mode. The top and bottom rows correspond to the ESPRIT and Cadzow algorithms, respectively, and the theoretical complex frequencies of the $(220)$ and $(320)$ QNMs are marked with open circles and squares. The extracted frequencies trace tight trajectories in the complex plane as a function of the analysis start time. The darker color shading corresponds to later start times (and therefore shorter ringdown signals). In the $(2,2)$ case (left), both algorithms show quick convergence toward the theoretical $(220)$ mode. In the $(3,2)$ case (right), the algorithms successfully separate the signal into its two distinct components: the native $(320)$ mode and the mixed-in $(220)$ mode, with both tracks tightly converging to their respective ground-truth values. We also attempted to extract a larger number of modes for each SXS channel, but observed that the quality of the recovery generally degraded. We therefore conclude that ESPRIT and the combined Cadzow-ESPRIT algorithms are most effective at simultaneously extracting a few of the loudest QNMs in a signal. In a complete data analysis pipeline, the loudest modes could be identified with ESPRIT and Cadzow, with subsequent subtraction and searches for subdominant modes using complementary techniques.

\section{Discussion\label{sec:discussion}}

\begin{table*}[htbp!]
    \centering
    \caption{Qualitative comparison of the three algorithms across the four test cases (Sections~\ref{subsec:single}--\ref{subsec:quasinorm}). In the first row, ``follows'' refers to the Fisher-optimal scaling $\mathcal{M}\propto\overline{\rho}\,{}^{-2}$ (see also Table~\ref{tab:fit_exponents}). In the frequency-separation experiment, all three algorithms resolve separations below the Fourier limit (``super-resolution''), with the errors $\sigma_f$ growing toward that limit as the SNR decreases. In the QNM experiment, ESPRIT is applied both directly and after Cadzow denoising, and frequency extraction always relies on ESPRIT.\label{tab:summary}}
    \vspace{0.5em}
    \setlength{\tabcolsep}{4pt}
    \renewcommand{\arraystretch}{1.5}
    \begin{tabular}{c||c|c|c}
        \hline\hline
         & ESPRIT & Cadzow & IRLS \\
        \hline
        \begin{tabular}{@{}c@{}}Mismatch scaling\\ ($\mathcal{M}\propto\overline{\rho}\,{}^{-2}$)\end{tabular} & \begin{tabular}{@{}c@{}}follows ($\rho\gtrsim 5$)\\ largest scatter\end{tabular} & \begin{tabular}{@{}c@{}}follows ($\rho\gtrsim 7$)\\ intermediate scatter\end{tabular} & \begin{tabular}{@{}c@{}}follows, small offset\\ smallest scatter\end{tabular} \\
        \hline
        \begin{tabular}{@{}c@{}}Number of components\\ ($n_{\rm true}$ unknown)\end{tabular} & \multicolumn{3}{c}{``elbow'' locates $n_{\rm true}$; very weak components missed} \\
        \hline
        \begin{tabular}{@{}c@{}}Frequency separation\\ (error $\sigma_f$)\end{tabular} & intermediate & lowest (esp. at low SNR) & highest; offset \\
        \hline
        \begin{tabular}{@{}c@{}}QNMs\\ (\texttt{SXS:BBH:0305})\end{tabular} & \multicolumn{2}{c|}{recovered (tight tracks)} & does not apply \\
        \hline\hline
    \end{tabular}
\end{table*}

In this work we tested three Hankel-based algorithms (ESPRIT, Cadzow iterations, and IRLS) for denoising time series containing single and multiple monochromatic (and possibly damped) signals. All three algorithms exhibit near-optimal performance consistent with the lower bounds on parameter uncertainties derived from a Fisher matrix analysis. This is quantitatively confirmed by the inverse-square scaling between the mismatch and the SNR. We also demonstrated the ability of these methods to estimate the number of components in a signal mixture by identifying an ``elbow'' in the mean-square residual as a function of the trial number of components. Finally, as a proof of concept, we demonstrated the algorithms' potential for recovering the frequencies and damping times of QNMs. 

Table~\ref{tab:summary} summarizes the comparative performance of the three algorithms across the four experiments. Cadzow iterations appear to provide the most robust results across our experiments and at varying levels of noise. While IRLS demonstrates the smallest scatter in the mismatch scaling, it exhibits a small but systematic offset in the single-signal and frequency separation cases, which we attribute to regularization. This offset is reminiscent of underfitting in machine learning, where the regularization term in step~4 of Algorithm~\ref{alg:HankIRLS} may become too large, pulling the solution away from the true minimum. The optimal tuning of the IRLS hyperparameters, such as the regularization parameter~$\lambda$ and its update schedule, requires further research.

In the context of black hole spectroscopy, our application of ESPRIT and Cadzow to a numerical relativity SXS waveform demonstrates their ability to extract the loudest QNMs. Namely, we successfully disentangled the native $(320)$ and mixed-in $(220)$ QNMs in the $(3,2)$ spherical harmonic contribution. However, we found that attempting to extract a larger number of modes simultaneously degrades the overall quality and stability of the recovery. We also explored a sequential extraction approach: after recovering the frequency of the dominant $(220)$ mode, we estimated its amplitude and phase via ordinary least squares (fitting for the linear coefficients $x=a\cos\phi$ and $y=a\sin\phi$) and subtracted it from the time series. This approach similarly resulted in poorer recovery of the remaining modes. Therefore, we conclude that in QNM spectroscopy, these Hankel-based techniques are best used to isolate the few loudest modes. They can serve as a first step in a hierarchical analysis pipeline, with subsequent searches for weaker modes using complementary techniques.

As GW detectors become more sensitive and the observation time increases, the resulting time series will grow significantly longer (for example, in LISA). Owing to the special structure of Hankel matrices, core operations, such as multiplication by a vector, can be implemented using the Fast Fourier Transform (FFT). This results in an efficient scaling of the Hankel-based algorithms with the length of the time series, $L$. The projection onto the Hankel subspace, for example, which involves averaging over anti-diagonals, is equivalent to a convolution. The continuous analogue of summing a function~$H(t, t')$ over its ``anti-diagonals'' $t+t'=\tau=\mbox{const}$\,\footnote{Here we assume that the $t$- and $t'$-axes in the $(t,t')$-plane follow the matrix convention, such that the positive directions are down (``rows'') and right (``columns''), respectively.} can be expressed as:
\beqa
    \label{eq:antidiag_conv}
    G(\tau) &=& \int\limits_{-\infty}^{+\infty}{H(t, \tau-t)\,\dd{t}}\,, \\
    \widetilde{G}_{f} &=& \widetilde{H}_{f,f}\,,
\eeqa
where $\widetilde{G}_{f}$ is the 1D Fourier transform of $G$, and $\widetilde{H}_{f,f'}$ is the 2D Fourier transform of $H$; the equality uses the diagonal slice $\widetilde{H}_{f,f}$. By the convolution theorem, the Fourier transform of the anti-diagonal averages corresponds to a slice of the 2D Fourier transform. This property allows the discrete operation to be implemented efficiently using 1D FFTs, which scale as $\mathcal{O}(L \log L)$. The SVD step represents another potential bottleneck. However, since only the top $r$ singular values are required, a truncated (or partial) SVD is sufficient. Combining fast FFT-enabled matrix-vector multiplications of Hankel matrices with the idea of computing only a truncated SVD, it is possible to compute step 1 of Algorithm~\ref{alg:esprit} or step 5 of Algorithm~\ref{alg:HankIRLS} in $\mathcal{O}(rL\log L)$ time~\cite{2009arXiv0911.4498K,Potts2015fast}, without the need to assemble a dense Hankel matrix as an intermediate step. Moreover, more sophisticated algorithms can achieve a sublinear runtime $\sim\mathcal{O}(\log L\log(1/\delta_\varepsilon))$~\cite{Kapralov-SODA:2026}, where $\delta_\varepsilon$ is a measure of the deviation between a Hankel matrix and its low-rank approximation.

We partially leveraged these properties to implement Cadzow iteration routines in TensorFlow~\cite{tensorflow2015-whitepaper,tensorflow2025-Zenodo}, which allow for efficient batch computations over multiple noise realizations simultaneously~\cite{lisahankel}. Although we used the full SVD (to our knowledge, currently there is no native TensorFlow implementation of partial SVD), we still achieved a significant speed-up compared to SciPy~\cite{Virtanen:2019joe} (which does have a truncated SVD routine). While we defer a detailed performance analysis to future work, typical runtimes for double precision are $\lesssim 1$~ms per timestep. This indicates that it is feasible to scale this denoising technique to much longer time series. In its current form, our implementation can already be applied to shorter, LIGO-scale time series such as the GW231123 event. 

Our implementation of IRLS, on the other hand, is based on NumPy~\cite{Harris:2020xlr} and SciPy~\cite{Virtanen:2019joe} and makes use of SciPy's \texttt{LinearOperator} class for computational efficiency. The complex logic of the weight operator in IRLS requires more work to implement in a GPU-friendly framework such as TensorFlow or PyTorch. The main computational bottleneck is the inversion of the linear operator associated to the weighted least squares problem in step~4 of Algorithm~\ref{alg:HankIRLS}. For the time-series length used in this paper ($L=400$), we found no significant performance difference between a direct dense matrix inversion and an iterative solver like BiCGSTAB~\cite{vanderVorst:1992eku}. However, for longer time series, the $\mathcal{O}(L^3)$ complexity of dense inversion becomes intractable, and iterative solvers offer a more efficient approach. One framework that can be used in the future is CoLA~\cite{2023arXiv230903060P}, which supports large-scale iterative algorithms.

The algorithms we have tested assume the signals to be monochromatic or damped sinusoids, whereas a generic GW source exhibits some degree of frequency drift. For LISA, the monochromatic approximation is expected to cover a large part of the science case. Galactic DWD binaries, the most abundant class of resolvable sources, are predominantly quasimonochromatic. Such sources constitute at least $50\%$ of the population in synthetic catalogs. In addition to GW emission, corrections to the frequency derivative~$\dot{f}$ can be caused by tidal interactions and mass transfer. However, these non-GW factors typically lead to negative~$\dot{f}$, making the sources more monochromatic~\cite{Strokov:2023ypy}. The approximation also works well for non-merging stellar-mass binary black holes~\cite{Kyutoku:2016ppx}.

For sources with an appreciable frequency drift and a sufficiently high SNR, the frequency range swept by the source can be divided into segments, with the signal treated as monochromatic within each segment. This is similar to the representation of chirping signals in spectrograms built from short Fourier transforms, an approach actively used in LISA data analysis~\cite{Tenorio:2025gci,Nobili:2026cnr}. Within each segment, a slowly growing amplitude can also be accommodated, since the signal model, Eq.~\eqref{eq:timeseries:model}, allows for complex frequencies. The segments belonging to the same source can then be identified in postprocessing, since their phases and amplitudes are correlated. While this introduces an additional postprocessing step, no specific dependence of the frequency derivative~$\dot{f}$ on~$f$ is assumed. The segmented approach can therefore be used, for example, in tests of general relativity or in searches for GW signals of unknown nature. We do not expect it to be useful for weak chirping signals, whose SNR within individual segments is too low. However, these low-SNR signals (e.g., stellar-mass black-hole binaries) are challenging for LISA data analysis regardless of the method~\cite{Moore:2019pke,Bandopadhyay:2024nys,Bandopadhyay:2025fyx}.

Our analysis has assumed signals embedded in stationary, white Gaussian noise. For stationary but nonwhite noise with a PSD~$S_{\rm n}(F)$, the simplest extension is to whiten the data, which is a common preprocessing step in GW data analysis, provided the PSD is known or can be reliably estimated. Whitening preserves the monochromatic character of the signals: it merely rescales the amplitude of each signal, $a_k\to a_k/\sqrt{S_{\rm n}(f_k)}$. The whitened data therefore fall within the regime in which the algorithms are rigorously applicable.

Alternatively, the noise PSD can be incorporated into the algorithms directly. In the time domain, stationary nonwhite noise has a Toeplitz\,\footnotemark{} covariance matrix~$\f{C}$. Accordingly, the Euclidean norm $\|\f{\hat{h}}-\f{h}\|_2^2 = (\f{\hat{h}}-\f{h})^\dagger(\f{\hat{h}}-\f{h})$ is replaced by its weighted counterpart, 
$$
(\f{\hat{h}}-\f{h})^\dagger\f{C}^{-1}(\f{\hat{h}}-\f{h})\,.
$$
This extension is straightforward for IRLS: the second term in Eq.~\eqref{eq:IRLS} is replaced by the weighted form, and each iteration still amounts to a linear solve. For Cadzow iterations, the matrix norm can also, in principle, be modified: weighted versions of the underlying low-rank approximation problem have been studied (see Section~4.1.1 of Ref.~\cite{Gillard:2023hankel}, as well as Ref.~\cite{Gillard:2016weighted}). However, the modification is nontrivial for a general $S_{\rm n}(F)$, because the SVD-based rank-truncation step retains its closed form only for special classes of weighted norms. It also turns out that the unweighted, ``vanilla'' version of the algorithms may correctly extract signals from correlated noise, as illustrated by the red-noise (autoregressive) example in Section~6.1 of Ref.~\cite{Gillard:2023hankel}, though how robustly this works for various noise spectra needs further research.

For nonstationary noise, the covariance matrix~$\f{C}$ is no longer Toeplitz but a general positive-definite matrix. It is this loss of structure that increases the computational complexity: multiplication by a Toeplitz matrix is a convolution and can be implemented with FFTs, whereas a general~$\f{C}$ offers no such shortcut, and the linear-operator implementation mentioned above loses its advantage. Nonstationary noise, however, is a nontrivial problem for LISA data analysis in general, regardless of the method. In the frequency domain, the noise covariance matrix is no longer diagonal, and a naive definition of a time-dependent spectrum, such as the Wigner--Ville spectrum, fails to remain positive~\cite{Cornish:2026tjt}. These issues can be addressed by the recently developed dynamic-spectrum framework of Refs.~\cite{Cornish:2025awt,Cornish:2026tjt}. Our algorithms, in contrast, operate in the time domain by construction and can be applied directly, with the noise model entering through the weighted formulation described above.\addtocounter{footnote}{-1}\footnotetext{A Toeplitz matrix, which is constant along each diagonal, provides a discrete version of convolution: the convolution of two discrete vectors is equivalent to the multiplication of one of them by a Toeplitz matrix built from the other.} We leave a study of these possibilities for future research.

In conclusion, we have shown that Hankel low-rank approximation methods offer a transparent and computationally efficient framework for denoising time-series data. These algorithms can be integrated in GW data analysis pipelines and could serve, for example, as a preprocessing step for identification of the approximate number of signals and their parameters, thus providing a starting point for Bayesian methods such as transdimensional MCMC~\cite{Cornish:2020dwh,Green:1995mxx}. As a final note, these algorithms can also be used for time series completion~\cite{Gillard:2023hankel}, and their application to the problem of data gaps in LISA is left for future research. 

\begin{acknowledgments}
This research was partially supported by a seed grant from the Space@Hopkins initiative at Johns Hopkins University.
N.G. and E.B. are supported by NSF Grants No.~AST-2307146, No.~PHY-2513337, No.~PHY-090003, and No.~PHY-20043, by NASA Grant No.~21-ATP21-0010, by John Templeton Foundation Grant No.~62840, by the Simons Foundation [MPS-SIP-00001698, E.B.], by the Simons Foundation International [SFI-MPS-BH-00012593-02], and by Italian Ministry of Foreign Affairs and International Cooperation Grant No.~PGR01167.
This work was carried out at the Advanced Research Computing at Hopkins (ARCH) core facility (\url{https://www.arch.jhu.edu/}), which is supported by the NSF Grant No.~OAC-1920103.
The authors acknowledge the computational resources provided by the WVU Research Computing Thorny Flat HPC cluster, partly funded by NSF~OAC-1726534.
\end{acknowledgments}

\section*{Data availability}

The data that support the findings of this article are openly available~\cite{lisahankel}.

\appendix

\section{Discrete Noise\label{app:noise}}

Here we summarize a prescription for generating mock discrete time series for stationary, Gaussian noise. Consider a white noise time series~$\epsilon_l$, $l=1,\ldots, L$\,, with timestep~$\Delta t$. The amplitudes~$\epsilon_l$ are i.i.d. and drawn from a normal distribution with zero mean and dispersion~$\sigma$, $\epsilon_l\sim\mathcal{N}\left(0;\sigma\right)$. The dispersion is, in general, a function of the timestep. Indeed, for a bigger timestep~$\Delta t'\gg\Delta t$, the new random samples $\epsilon^\prime_m$ can be interpreted as averages over all~$\epsilon_l$ that are contained within the bigger interval~$\Delta t'$,
\beqa
\epsilon^\prime_m &=& \frac{\Delta t}{\Delta t'}\sum \epsilon_l\,, \\
\left\langle \epsilon^\prime_m\right\rangle &=& 0\,, \\
(\sigma')^2 &=& \left(\frac{\Delta t}{\Delta t'}\right)^2\sum \left\langle \epsilon_l^2\right\rangle = \sigma^2\frac{\Delta t}{\Delta t'}\,,
\eeqa
whence
\be
\label{app:eq:sigma_dt}
\sigma \propto \frac{1}{\sqrt{\Delta t}}\,.
\ee

In the Fourier domain, the respective amplitudes are $\widetilde{\epsilon}_{F_k}=\widetilde{\epsilon}_{k}\Delta t$, where
\beqa
F_k &=& k/(L\Delta t)\,, \quad  k=0,\ldots,L-1\,, \\
\widetilde{\epsilon}_k &=& \sum\limits_{l=1}^{L}\epsilon_l e^{-2\pi i k(l-1)/L}\,, \label{app:eq:DFT}
\eeqa
are correspondingly the discrete Fourier frequencies and the definition of the discrete Fourier transform (DFT) consistent with NumPy and SciPy conventions~\cite{Harris:2020xlr,Virtanen:2019joe}. The Fourier amplitudes are also i.i.d. and normally distributed with a dispersion that is related to the (one-sided) PSD \mbox{$S_{\rm n} = \mbox{const}$}:
\be
\left\langle \widetilde{\epsilon}_F \widetilde{\epsilon}{}^\star_{F'}\right\rangle = \frac 12 S_{\rm n} \delta(F-F')\,.
\ee
Therefore,
\be
\left\langle \left|\widetilde{\epsilon}_k\right|^2\right\rangle (\Delta t)^2 \Delta f = \frac 12 S_{\rm n} \quad \Rightarrow \quad \left\langle \left|\widetilde{\epsilon}_k\right|^2\right\rangle  = \frac{S_{\rm n}}{2(\Delta t)^2 \Delta f}\,,
\ee
where the additional factor~$\Delta f\equiv 1/(L\Delta t)$ comes from integration over a frequency bin.

On the other hand, using Eq.~\eqref{app:eq:DFT}, we obtain
\beqa
\left\langle \left|\widetilde{\epsilon}_k\right|^2\right\rangle &=& L\sigma^2\,, \\
\sigma^2 &=& \frac{S_{\rm n}}{2(\Delta t)^2 L\Delta f} = \frac 12 \frac{S_{\rm n}\Delta t}{(\Delta t)^2} = \frac{S_{\rm n}}{2\Delta t}\,,
\eeqa
which allows us to determine the proportionality factor in Eq.~\eqref{app:eq:sigma_dt}:
\be
\sigma = \sqrt{\frac{S_{\rm n}}{2\Delta t}}\,.
\ee
If the white noise is normalized such that $S_{\rm n}=1$, then simply $\sigma = 1/\sqrt{2\Delta t}$.

In particular, this leads to an algorithm for simulating time-domain noise with an arbitrary PSD $S_{\rm n}=S_{\rm n}(F)$:
\begin{itemize}
  \item draw $\epsilon_l$ from the normal distribution with a zero mean and $\sigma_{\Delta t} = 1/\sqrt{2\Delta t}$\,,
  \item run DFT to obtain the Fourier domain values~$\widetilde{\epsilon}_k$\,,
  \item scale the Fourier values, \mbox{$\widetilde{\epsilon}_k \rightarrow \widetilde{\epsilon}_k\sqrt{S_{\rm n}(F_k)}$}\,,
  \item run an inverse DFT\,.
\end{itemize}

\bibliography{references}

\end{document}